\documentclass{aa}  

\usepackage{graphicx}
\usepackage{txfonts}
\usepackage{lipsum}
\usepackage{subcaption}        
\usepackage{lscape}             
\usepackage{placeins}           
\usepackage{xcolor} 
\usepackage{natbib}
\usepackage{ulem} 
\usepackage{wasysym}
\usepackage{float}

\usepackage{hyperref}   
\hypersetup{
    colorlinks=true,
    linkcolor=blue,
    citecolor=blue,
    urlcolor=blue
}
\usepackage{indentfirst}
\usepackage{enumitem}

\newcommand{\cc}[1]{\textcolor{black}{#1}}

\begin{document}

   \title{White dwarf pollution in wide binaries I: the effects of stellar evolution}

   \author{C. Norambuena\inst{1}\fnmsep\thanks{constanza.norambuena@uc.cl},
           C. Charalambous\inst{1}
           \and C. Aguilera-Gómez\inst{1}
        }

   \institute{Instituto de Astrofísica, Pontificia Universidad Católica de Chile, Av. Vicuña Mackenna 4860, 782-0436 Macul, Santiago, Chile}

   \date{Received August 12, 2026}

  \abstract 
    {Atmospheric pollution in white dwarfs is interpreted as evidence for the accretion of rocky material from remnant planetary systems. \cc{Wide binary companions can influence the delivery of planetary material onto white dwarfs, making it essential to understand the interplay between stellar evolution and binary perturbations.}}
    {We investigate how uncertainties in stellar mass-loss histories affect the dynamical evolution of remnant planetary systems in wide binaries, and assess their relative importance compared with binary perturbations in driving the accretion of rocky material onto white dwarfs.} 
    {We perform \texttt{REBOUND} N-body simulations incorporating time-dependent stellar parameters interpolated from \texttt{MESA} evolutionary tracks and tidal interactions, following the dynamical evolution up to the white dwarf stage. We explore two stellar-evolution models with distinct mass-loss histories for wide binary systems hosting a Jupiter-mass planet and a planetesimal disc, spanning a range of binary architectures.}
    {The adopted stellar-evolution model has only a minor effect on the survival of the planet, mainly affecting the timing and efficiency of planetesimal delivery during the white dwarf stage. Instead, the architecture of the binary primarily controls the pollution outcome: wider binaries produce the highest rates of accretion onto the white dwarf, consistent with the observed prevalence of polluted white dwarfs in wide binaries. Moreover, planetesimal reservoirs remain long-lived, with survival times increasing with binary separation, enabling Gyr-long metal pollution.}
    {Our results suggest that the architecture of wide binaries, rather than uncertainties in stellar evolution, is the dominant factor controlling the dynamical delivery of rocky material onto white dwarfs.}


\titlerunning{White dwarf pollution in wide binaries}
    \maketitle
    \nolinenumbers
\section{Introduction}

White dwarfs (WDs) are the final evolutionary stage of approximately $95\%$ of all stars in the Milky Way. Their atmospheres are expected to be composed almost entirely of hydrogen or helium, as heavy elements rapidly sink below the photosphere due to the strong surface gravity and weak radiative forces \citep{1995Chayer, 2009Koester}. However, \cc{$25\%-50\%$ of single WDs exhibit metal absorption lines in their atmospheres \citep{2014Koester, 2024OuldRouis}, although the reported fraction depends on observational and detection criteria.} These objects are referred to as polluted WDs.  

The composition of the detected metals in these polluted WDs is broadly consistent with rocky material \citep{2007Zuckerman, 2012Gansicke, 2013Farihi, 2014Jura, 2014Xu}, suggesting that the pollution originates from minor bodies\footnote{Throughout this work, the term planetesimal is used in a broad sense to denote minor bodies that can act as sources of WD pollution.}, such as asteroids, comets, satellites, or rocky debris, that survived the post-main-sequence (post-MS) evolution of the planetary system, although the tidal disruption of larger planetary bodies may also contribute. 

These bodies can be perturbed onto star-grazing orbits, become tidally disrupted by the WD, and subsequently accreted onto its surface \citep{2016Farihi}. Observational evidence supports this scenario, including the detection of circumstellar debris discs through infrared excess around many polluted WDs (e.g., \citealt{2024Rogers,2024MadurgaFavieres}, and see also the review by \citealt{2016Farihi}), \cc{and the direct observation of disintegrating planetary material transiting WDs \citep[e.g.,][]{2015Vanderburg, 2020Vanderbosch, 2025Guidry, 2025Bhattacharjee}.} 

Given that planetary systems and debris discs are common around MS stars \citep{2012Cassan, 2013Eiroa,2021Zhu}, a fraction of these rocky bodies is expected to survive the post-MS evolution and remain present during the WD stage \cc{\citep{2018Mustill, 2021Smallwood}}. Whether such remnant planetary systems survive, however, depends sensitively on the post-MS evolution of the host star. The long-term survival of these remnant planetary systems is governed by the competition between stellar mass-loss and tidal interactions during the giant phases \citep{2016Veras}. While tides tend to shrink planetary orbits and promote engulfment, stellar mass-loss drives orbital expansion. Their interplay ultimately determines the architecture of the surviving planetary system and the availability of planets and planetesimal reservoirs capable of delivering rocky material onto the WD.

Mass-loss during the red giant branch (RGB) and the asymptotic giant branch (AGB) phases governs the removal of the stellar envelope and the transition toward the WD stage \citep{2018Hofner}. Despite decades of observational and theoretical efforts, its efficiency and dependence on stellar properties remain uncertain \citep{2015Mcdonald, 2018Hofner}. Although empirical prescriptions are widely adopted in stellar-evolution calculations, the physical mechanisms driving stellar winds, particularly during the AGB phase, where pulsation-enhanced and dust-driven outflows dominate, are still not fully understood \citep{2015Hofner,2018Hofner}. Consequently, even modest differences in the adopted mass-loss prescription may lead to different planetary architectures and alter the efficiency with which remnant rocky material is delivered onto the WD.

Previous studies show that stellar evolution can significantly modify the architecture of planetary systems and trigger dynamical instabilities capable of delivering rocky material onto WDs. \cc{Numerical simulations have demonstrated that stellar mass-loss can destabilize remnant planetesimal belts \citep{2011Bonsor}, while the survival of planetary systems through the RGB and AGB phases has been also investigated \citep{2012Mustill,2020Ronco}.} Together, these studies highlight the key role of post-MS stellar evolution in shaping planetary architectures that may lead to WD pollution.

\begin{figure}
    \centering
    \includegraphics[width=\hsize]{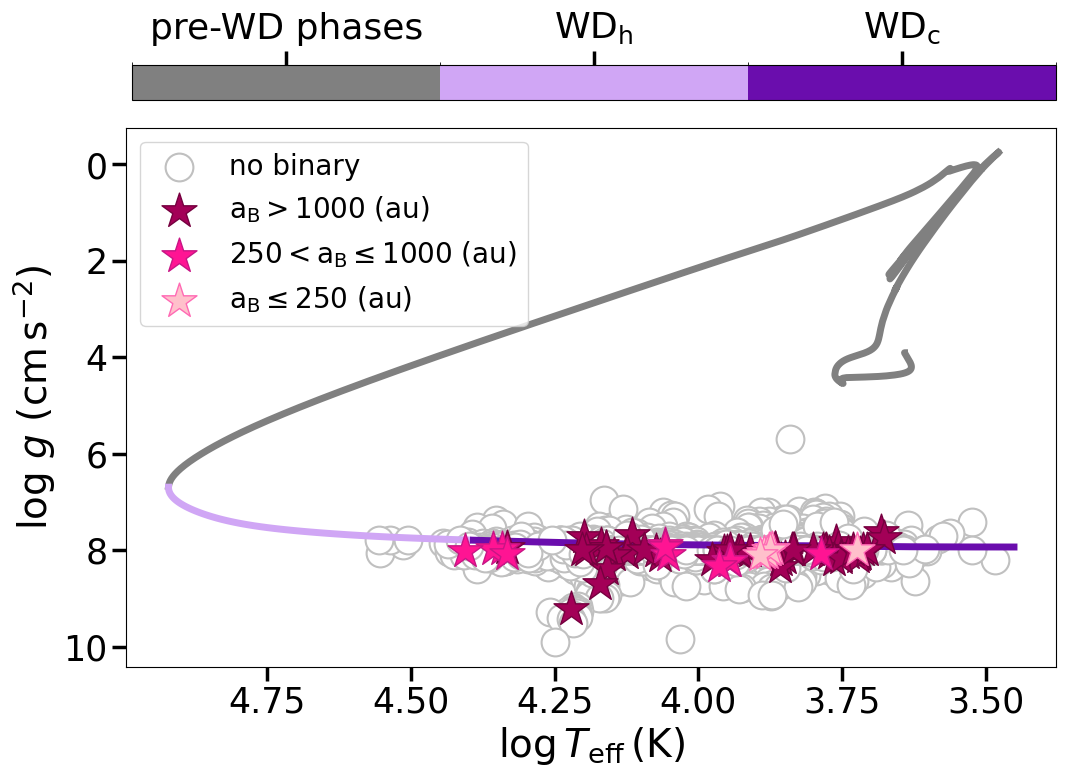}
    \caption{Evolutionary track of a Sun-like star\cc{, showing the evolution of its effective temperature and surface gravity from the pre-WD phase (gray) through the WD stage.} The hotter ($T_{\mathrm{eff}} \geq 25,000,\mathrm{K}$) and cooler ($T_{\mathrm{eff}} < 25,000,\mathrm{K}$) WD phases are shown in light and dark purple, respectively. Polluted WDs from the PEWDD catalog \citep{2024Williams} are overplotted as white circles for single systems and star symbols for binary systems, with the symbol colours indicating the binary separation.}
    
\label{fig:evol_track_observations}
\end{figure}

Fig.~\ref{fig:evol_track_observations} shows the distribution of polluted WDs from the Planetary Enriched White Dwarf Database (PEWDD; \citealt{2024Williams}) in the Kiel diagram\cc{, which relates stellar effective temperature to surface gravity}. Empty circles denote single polluted WDs, whereas star symbols denote polluted WDs in binary systems, with colours indicating the binary projected separation. The evolutionary track of a Sun-like star is overplotted to highlight the different evolutionary stages considered throughout this work. The track is shown in gray during the pre-WD evolution and in purple after WD formation. 

We distinguish between a hotter WD phase (WD$_{\rm h}$; light purple), with $T_{\rm eff}\geq25,000$~K, and a cooler WD phase (WD$_{\rm c}$; dark purple), with $T_{\rm eff}<25,000$~K. This distinction is motivated by the fact that radiative levitation can support heavy elements in the atmospheres of hotter WDs, making the origin of the detected metals more ambiguous. In contrast, for WDs with temperatures below $25,000$~K, radiative levitation becomes inefficient, and the presence of metals is generally interpreted as evidence for ongoing or recent accretion of planetary material \citep{2016Farihi, 2009Koester, 1995Chayer}. \cc{We therefore use WD$_{\rm h}$ and WD$_{\rm c}$ as observationally motivated categories to distinguish temperature regimes in which atmospheric metal pollution has different physical interpretations. In particular, the cooler WD$_{\rm c}$ regime provides a cleaner observational context in which detected metals can be interpreted as evidence of ongoing or recent accretion of planetary material.}

\cc{Although a fraction of polluted WDs are found in wide binary systems, most are not known to have stellar companions. The observed binary fraction depends strongly on sample selection: while some samples of polluted WDs have binary fractions comparable to that of the field, the most strongly polluted samples show substantially lower binary fractions, with fractions of about 2\% reported for the most strongly enriched populations \citep{2019Wilson,2024Noor, 2026Xu}. In addition, no infrared-dusty polluted WDs are currently known to have stellar companions, apart from the circumbinary disc associated with the post-common-envelope binary SDSS 1557 \citep{2019Wilson}. These observations suggest that the relationship between binarity and atmospheric pollution is not simply a matter of whether a WD has a companion, but may depend on the binary architecture and on the properties of the polluted WD population. Wide stellar companions can induce secular perturbations that either enhance or suppress the delivery of planetesimals onto the WD, depending on the binary configuration \citep{2018Veras,2026AvilaBravo}. Understanding these effects is therefore important for interpreting the occurrence and properties of polluted WDs in binary systems.}

The dynamical role of wide stellar companions in WD pollution has been investigated in several previous studies. \cc{Semi-analytic studies have shown that the engulfment of an inner planet during the AGB phase can remove the dynamical shielding of an exterior planetesimal disc, allowing von Zeipel-Kozai-Lidov (ZKL) oscillations induced by a wide stellar companion during the WD phase and promoting the tidal disruption of rocky bodies \citep{2017Petrovich}. Post-MS stellar evolution in wide binary systems can also trigger the eccentric ZKL mechanism through stellar mass loss, driving planets and minor bodies onto highly eccentric orbits that may ultimately pollute the WD \citep{2017Stephan}.}

\cc{Direct N-body studies have further demonstrated that stellar mass loss can destabilize planetary systems that were initially stable, with the resulting stability boundary depending strongly on the binary architecture \citep{2017Veras}. Galactic tides can additionally perturb wide binary orbits and destabilize remnant planetary systems during the WD phase \citep{2015Bonsor}.} Together, these studies have established the importance of stellar evolution and wide binary companions in shaping the long-term dynamics of planetary systems. However, the combined influence of stellar mass-loss histories, tidal interactions, and wide-binary perturbations on the evolution of planetesimal reservoirs has received less attention.

In this work, we perform N-body simulations of wide binary systems in which the host star evolves into a WD while hosting a giant planet and a planetesimal disc. We explore a broad range of binary configurations and planetary architectures, incorporating stellar mass loss, tidal interactions throughout the giant phases, and two stellar-evolution models with different mass-loss histories. This allows us to investigate how the coupled effects of stellar evolution and binary perturbations shape the survival of planetary system components and promote the accretion of planetesimals onto the WD, potentially leading to atmospheric pollution. In Sect.~\ref{sec:stel_evol} we describe the stellar-evolution models adopted in this work. In Sect.~\ref{sec:nbody_dynamics} we present our simulation setup, initial conditions, and the implementation of stellar evolution and tidal effects within the N-body simulations. In Sect.~\ref{sec:results} we present and analyze the results of our parameter exploration. We discuss the implications of these results in Sect.~\ref{sec:discussion} and summarize our main findings in Sect.~\ref{sec:conclusion}.

\section{Methods}\label{sec:Methods}

Each system consists of a central star and a binary companion. The binary companion is assumed to be less massive than the host star and therefore has a longer evolutionary timescale, remaining essentially unevolved throughout the duration of our simulations. Consequently, only the central star undergoes stellar evolution, while the binary companion acts as a gravitational perturber.

To model these systems, we combine stellar-evolution calculations for the central star with direct N-body integrations. Starting from the post-RGB phase, we follow the dynamical evolution of planetary systems as the host star evolves into a WD. 

    \subsection{Stellar evolution} \label{sec:stel_evol}

    We use the stellar-evolution code \texttt{MESA} (Modules for Experiments in Stellar Astrophysics, \citealp{Paxton2011, Paxton2013, Paxton2015, Paxton2018, Paxton2019, Jermyn2023}, version 24.07.1) to compute the evolution of the host star from the zero-age main sequence (ZAMS) to the WD phase. 
    By solving the 1D stellar structure and evolution equations, \texttt{MESA} provides the full stellar-evolutionary track, from which we extract the time-dependent stellar mass $M(t)$, radius $R(t)$, and luminosity $L(t)$, which are then incorporated into our N-body simulations. We adopt the default physical prescriptions and numerical settings of the \texttt{MESA} version except for the parameters explicitly listed in Appendix~\ref{appendix:methodology}.
    
    \begin{figure}
        \centering
        \includegraphics[width=\hsize]{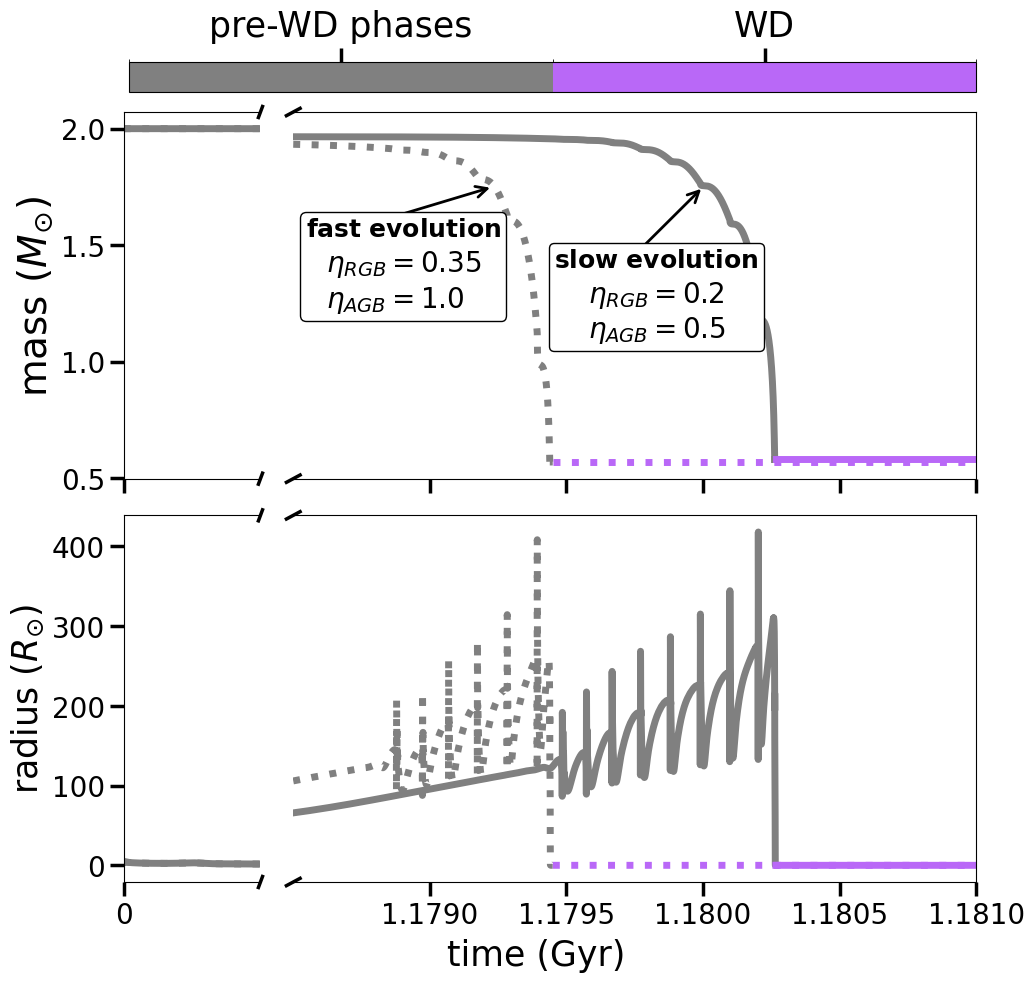}
        \caption{Evolution of the stellar mass (top) and radius (bottom) for a host star with an initial mass of $M_{\rm MS} = 2\,M_{\odot}$ and metallicity $Z = 0.02$. \cc{The time axis corresponds to the stellar age, with a short segment of the main sequence evolution and the later transition from the AGB to the WD phase. The colour bar indicates the evolutionary stage of the host star: the pre-WD phases are shown in gray, while the WD phase is shown in purple.}The dotted line corresponds to the fast evolution model, while the solid line shows the slow evolution track.}
        \label{fig:mass_radius_evolution}
    \end{figure}
    
    We consider a host star with an initial mass in the main sequence of $M_{\rm MS} = 2 \,M_{\odot}$ and metallicity $Z = 0.02$. Mass-loss along the RGB is modeled following the Reimers prescription \citep{1975Reimers}, parameterized by an efficiency factor $\eta_{\mathrm{RGB}}$, while mass-loss during the AGB follows the Blöcker prescription \citep{1995Blocker}, with an efficiency parameter $\eta_{\mathrm{AGB}}$.

    The WD cooling sequence is evolved using the standard \texttt{MESA} cooling treatment, without including additional physical processes such as crystallization or sedimentation-induced heating associated with element diffusion. The WD is evolved until its luminosity decreases below $L <10^{-3}L_{\odot}$.

    Because mass-loss remains an open topic in stellar astrophysics \citep{2018Hofner, 2015Mcdonald}, we explore two configurations for post-MS phases: a fast evolution model and a slow evolution model. This allows us to assess how different stellar evolutionary timescales affect the system dynamics. The resulting evolution of stellar mass and radius near the WD phase is shown in Fig.~\ref{fig:mass_radius_evolution}. In both scenarios, the star reaches a final WD mass of $M_{\rm WD}= 0.6\, M_\odot$, although the mass-loss histories differ.
    
    The characteristic thermal pulses of the AGB phase are clearly visible in both models.
    Although the fast evolution model exhibits fewer pulses, the main difference is that it reaches the WD phase slightly earlier than the slow evolution model, while both reach nearly the same maximum stellar radius. We note, however, that even a small difference in the timing of the mass-loss phase from the perspective of stellar evolution ($\sim 0.01$ Gyr) becomes dynamically important for the planetary system, as it corresponds to many orbital periods of the planets and planetesimals and can therefore lead to significantly different dynamical outcomes, ultimately affecting the efficiency of WD pollution.

    If only stellar mass-loss is considered, orbits expand adiabatically as the stellar mass decreases. However, this effect alone does not provide a complete description of the orbital evolution. Tidal interactions can counteract this expansion and substantially alter the orbital evolution of close-in planets and planetesimals.
        
    Stellar tides arise from the gravitational distortion induced by a planet on its host star, with the associated dissipation transferring angular momentum between the stellar spin and the orbit, leading to orbital decay and eccentricity damping for close-in planets. During post-MS evolution, these effects become increasingly important as the stellar radius expands, enhancing tidal dissipation and potentially driving planets and planetesimals toward engulfment. Since the characteristic tidal timescales depend strongly on the stellar mass and radius, even modest differences in the stellar-evolution tracks can lead to substantially different orbital evolutions. In particular, during the giant phases, the rapid increase in stellar radius greatly enhances tidal dissipation, making the detailed temporal evolution of the star important for the fate of nearby planets and planetesimals (Further details can be found in Appendix~\ref{appendix:methodology}).

\subsection{N-body dynamics}\label{sec:nbody_dynamics}
\begin{table}
    \centering
    \begin{tabular}{cccc}
        \hline
        &  Planet & Planetesimals & Binary\\
        \hline
        Mass & $1M_{\mathrm{J}}$ & $4M_{\leftmoon}$ & $1M_{\odot}$\\
        a (au)& 2, 3, 5 & inner, outer & 250, 500, 1000\\
        e & [$10^{-4}$, $10^{-3}$] & [$10^{-4}$, $10^{-3}$] & 0.01, 0.3, 0.6, 0.9\\
        i (deg) & [-1, 1] & [-1, 1] & 0.01, 30, 60, 90\\
        \hline
    \end{tabular}
    \caption{Initial conditions used in the N-body simulations, including the masses and orbital parameters of planet, planetesimals, and binary companion.}
    \label{tab:parameters}
\end{table}
        
We model the late-stage dynamical evolution of planetary systems by coupling direct N-body integrations with stellar-evolution tracks derived from \texttt{MESA}. The dynamical evolution is computed with the open-source code \texttt{REBOUND} \citep{2012ReinLiu}, while the stellar-evolution models provide the time-dependent stellar properties required throughout the integrations.
    
As described in Sect.~\ref{sec:stel_evol}, \texttt{MESA} provides the temporal evolution of the stellar mass, radius and luminosity, which are used to compute the $\tau$ parameter through Equation~\ref{eq:tau}. Because dynamical timescales are much shorter than those over which stellar properties evolve, recomputing stellar quantities at each integration step would be computationally prohibitive. Stellar evolution and stellar tidal effects are therefore implemented through the \texttt{REBOUNDx} library \citep{2020Tamayo}, together with its parameter interpolation module \citep{2022Baronett}, which evolves $M(t)$, $R(t)$ and $\tau(t)$ during the N-body integration.

In our numerical simulations, we focus exclusively on the evolutionary phases from the horizontal branch (HB) onward. This choice is motivated both by computational cost and by our specific interest in the late stages of stellar evolution, where mass-loss and orbital expansion have the strongest impact on the system dynamics. We therefore implicitly assume that the planetary systems considered are those that survive the earlier post-MS evolution prior to the HB phase.

The initial setup consists of a host star orbited by a Jupiter-like planet and a planetesimal disc, together with a $1\,M_{\odot}$ stellar companion acting as an external perturber. Table~\ref{tab:parameters} summarizes the adopted initial conditions. Throughout this work, quantities associated with the binary companion are denoted with the subscript B \cc{(e.g., semi-major axis $a_{\rm B}$, eccentricity $e_{\rm B}$, and inclination $i_{\rm B}$)}, those corresponding to the planet with the subscript p \cc{($a_{\rm p}$, $e_{\rm p}$, and $i_{\rm p}$)}, while planetesimal quantities are written without subscripts \cc{($a$, $e$, and $i$)} unless otherwise specified.

Observational constraints on binary orbital parameters remain limited, as in most cases only the projected separation is known \citep{2018ElBadry}. Motivated by these uncertainties, as well as by the observed distribution of polluted WDs shown in Fig.~\ref{fig:evol_track_observations}, we explored a broad range of binary orbital parameters, including different binary eccentricities, inclinations, and semi-major axes. 

Wide binaries ($a_{\rm B}\gtrsim250$~au) evolve nearly independently from the perspective of stellar evolution, largely avoiding strong binary interactions such as mass-transfer \citep{2018Hogg} or common envelopes \citep{2010NewAR..54...65T,2013A&ARv..21...59I,2013MNRAS.436.2515M}. Furthermore, sufficiently close stellar companions may themselves become the dominant source of WD pollution rather than the planetary system \citep{2018Veras}. To avoid these effects and to focus on pollution driven by the planetary system, we adopt $a_{\rm B}=250$~au as the lower limit for the binary separation. At sufficiently large separations, Galactic tides are expected to play an increasingly important role in the dynamical evolution of the system \citep{2013Kaib}. We therefore adopt $a_{\rm B}=1000$~au as an upper limit, leaving the exploration of wider binaries for future work.

To explore the different dynamical regimes accessible within this parameter space, we consider a grid spanning inclinations between $i_{\mathrm{B}}=0^{\circ} - 90^{\circ}$, eccentricities from $e_{\mathrm{B}} =0 - 0.9$ and semi-major axis in the range $a_{\mathrm{B}}=250 - 1000$~au. This range is chosen to probe different dynamical regimes: mutually inclined configurations ($i_{\rm B} \geq 40^\circ$) can trigger von Zeipel-Kozai–Lidov (ZKL) oscillations \citep{1910vonZeipel,1962Kozai,1962Lidov}, while high binary eccentricities favour the eccentric Kozai–Lidov mechanism \citep{2013Naoz}. At wider separations ($a_{\rm B} \gtrsim 1000$~au), external perturbations such as Galactic tides may also become relevant \citep{2013Kaib}. Since our grid extends up to $a_{\rm B}= 1000$~au, we do not include the effects of the Galactic potential, which are expected to become important mainly at larger binary separations \citep{2015Bonsor, 2023Modak}, and will be addressed in future work.  

We explore different system architectures, considering, in separate simulations, a planetesimal disc located either interior or exterior to the planetary orbit. The planet is initially placed  $a_{\mathrm{p}}$=2, 3\cc{,} or 5~au on a quasi-circular orbit, with the disc location shifted accordingly in each case. The physical motivation for the adopted parameters can be found in Appendix~\ref{appendix:methodology}.

In the following, we \cc{primarily} focus on the configuration with $a_{\rm p}=3$~au, which we adopt as our fiducial case \cc{and discuss in detail in Sect.~\ref{sect:fiducial_case}}. In this setup, the inner planetesimal disc extends over $a \in [1.25,2.5]$~au, while the outer disc spans $a \in [3.5,5]$~au. The configurations with \cc{the planet located at} $a_{\rm p}=2$ and 5~au are discussed in \cc{Sect.~\ref{sect:2_5au}.}

\cc{
The planetesimal disc is initialized at the HB with 25 bodies, each with a mass of $4\,M_{\leftmoon}$, corresponding to a total initial disc mass of $100\,M_{\leftmoon}$ ($\simeq1.2\,M_{\oplus}$). This mass is substantially larger than the estimated masses of both the present-day Solar System Kuiper Belt ($\simeq0.02\,M_{\oplus}$; \citealt{2018Pitjeva}) and Main Belt ($\simeq0.0005\,M_{\oplus}$; \citealt{2018Pitjeva}), and should therefore not be interpreted as a direct analogue of either belt. Instead, it represents an idealized reservoir of rocky material at the onset of post-main-sequence evolution, designed to investigate the dynamical pathways through which material can be delivered onto the WD during stellar evolution. More massive debris discs are observed around other stars; for example, the debris disc around Vega ($2.3\,M_\odot$) has been estimated to contain of order $10\,M_\oplus$ in $\sim$100-km-sized planetesimals \citep{2010ApJ...708.1728M}. Our adopted disc therefore represents a relatively massive reservoir of solid material, rather than a direct analogue of either Solar System belt.}

\cc{Real planetesimal and debris discs contain much larger populations spanning a broad range of sizes and masses, and their evolution is affected by collisions, fragmentation, and the resulting mass redistribution \citep{2018Hughes, 2021Krivov}. These processes are not included in our simulations. Instead, we represent the reservoir with 25 gravitationally interacting bodies, which provides a computationally tractable approximation while allowing us to follow the dynamical evolution and mutual interactions of the solid material. This simplification may enhance the importance of mutual scattering relative to a realistic population with a broad size distribution, and the resulting depletion and delivery timescales should therefore not be interpreted as direct predictions for a collisionally evolved debris disc. Our simulations should instead be understood as exploring the dynamical evolution of a reservoir of relatively massive rocky bodies during stellar evolution; modelling its detailed size distribution and collisional evolution is beyond the scope of this work. The adopted disc mass can also be placed in the broader context of studies of long-lived planetesimal discs, which have investigated the masses and planetesimal properties required for dynamical survival over Gyr timescales \citep{2010MNRAS.401..867H}. However, these studies do not account for stellar evolution and the associated changes in the gravitational potential. Thus, although a planetesimal reservoir of this mass may remain dynamically stable over Gyr timescales around a main-sequence star, its survival cannot be assumed once the host star evolves off the main sequence and undergoes substantial mass loss. Assessing the subsequent dynamical evolution of such a reservoir during stellar evolution is precisely the focus of this work.}

Because the system contains many interacting bodies, close encounters can frequently occur. In addition, the inclusion of stellar evolution and tidal effects makes the simulations computationally expensive. For these reasons, we employ the \texttt{TRACE} hybrid integrator \citep{2024Lu}, which combines the high accuracy and adaptive time-stepping of the \texttt{IAS15} integrator \citep{2015ReinSpiegel} during close encounters within three Hill radii ($3\,R_{\mathrm{H}}$), with the computational efficiency of the \texttt{WHFAST} integrator \citep{2015ReinTamayo} otherwise, making it well suited for hierarchically structured systems with an external stellar perturber.

In total, we performed 1344 simulations spanning the full range of binary, planetary, and planetesimal disc configurations described above. Five independent simulations were performed for each fiducial configuration, while only one simulation per configuration was carried out for the remaining planetary orbital distance.

\section{Results}\label{sec:results}

The simulations start at the end of the HB phase and follow through the entire AGB phase, which we refer to as the ``pre-WD phases''. During these stages, some planetesimals undergo significant dynamical perturbations. Some are engulfed by the expanding host star after reaching a critical radius, while others are ejected from the system, characterized by highly eccentric orbits \cc{($e \approx 1$)} and very large pericenter distances, indicating that they become gravitationally unbound.
In this work, we do not model the long-term MS evolution of the system, and instead assume that the initial architecture remains stable until the onset of post-MS evolution. Our focus is therefore on the subsequent multi-body dynamical evolution during the late stages of stellar evolution and the WD phase, where stellar mass-loss and radius expansion can strongly reshape the system \citep{2012MNRAS.420.2990B}.

At the onset of the WD phase, when the temperature exceeds \cc{$\approx$25,000~K}, radiative levitation can support metals in the stellar atmosphere \citep{2016Farihi, 1995Chayer,2009Koester}, making it difficult to distinguish between internally supported metals and those delivered through external accretion. When the temperature falls below this threshold and the observed metal pollution can be more reliably attributed to the accretion of rocky material from a debris disc. We consider the end of the relevant phase to occur when the WD cools down to $T_{\rm eff}$=5,000~K. We adopt this threshold because cooler WDs are intrinsically faint, introducing observational biases that hinder their characterization through high-resolution spectroscopy \citep{2017Hollands, 2022Elms}. \cc{Thus, the distinction between WD$_{\rm h}$ and WD$_{\rm c}$ is observationally motivated and does not represent a physical transition within the WD.}

With these considerations in mind, our simulations span this entire time interval, corresponding to a total evolution time of \cc{$\approx 5$~Gyr}, with only minor differences of order \cc{$\approx 225$~Myr} between the fast and slow evolution models. This choice provides a realistic evolutionary timescale while maintaining a reasonable computational cost. In practice, some simulations terminate earlier when both the planet and the planetesimal disc are removed, either through ejection or accretion onto the central star, leaving only the host star and the binary companion in the system. 

\subsection{\cc{Fiducial case: $a_{\rm p}=3$~au}}
\label{sect:fiducial_case}
\begin{figure}
\centering
\includegraphics[width=\hsize]{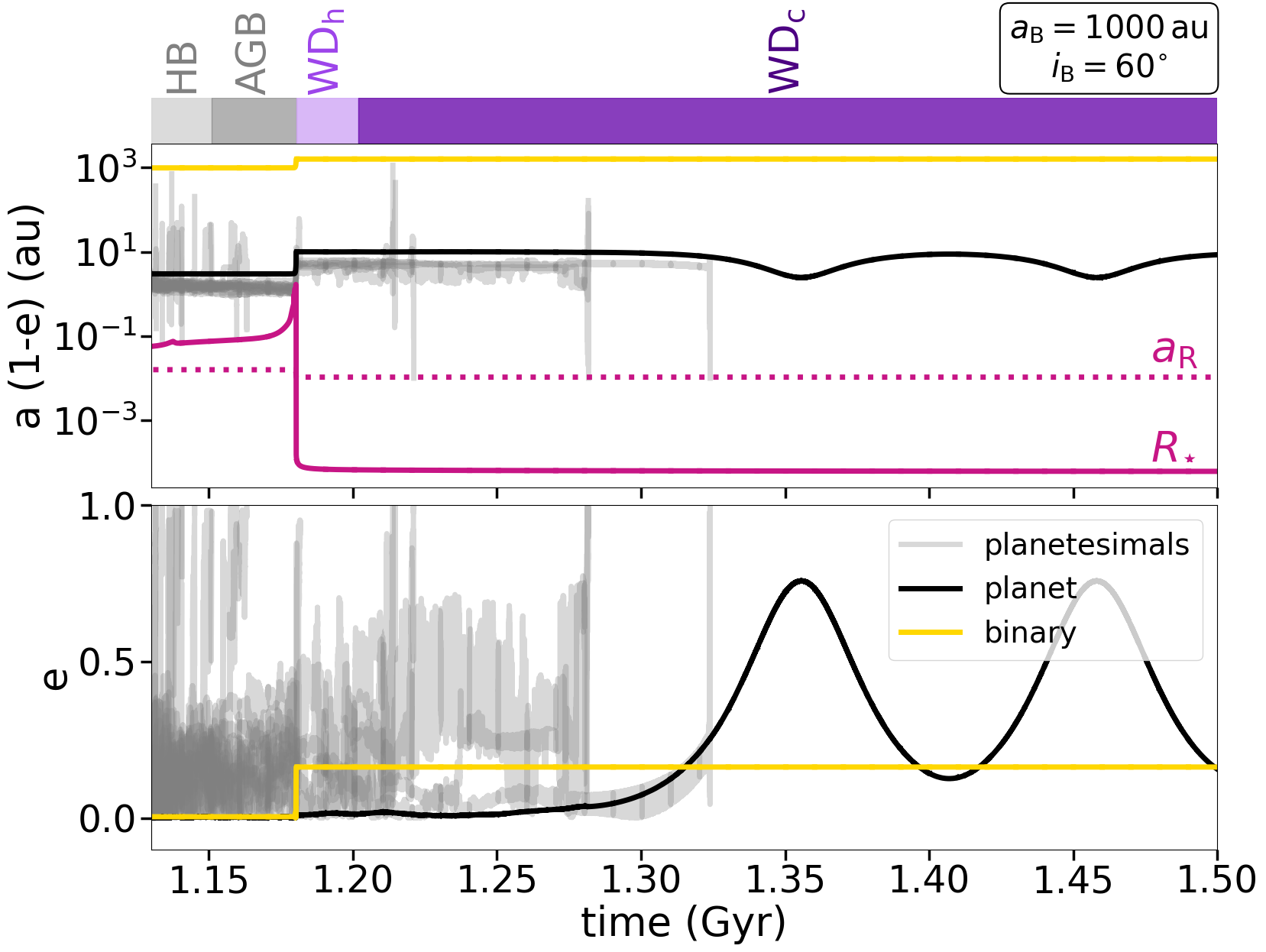}
    \caption{Temporal evolution of the pericenter (top) and eccentricity (bottom) of all bodies in a system with a slow evolving host star\cc{, with time corresponding to the host star's age, counting from the ZAMS}. The initial binary configuration has $a_{\mathrm{B}} = 1000$~au, $i_{\mathrm{B}} = 60^\circ$, and $e_{\mathrm{B}} = 0$ (yellow). The planet (black) is located at $a_{\mathrm{p}} = 3$~au, while the planetesimal disc (gray) is placed interior to it, spanning $a \in [1.25, 2.5]$~au. The colour bar indicates the evolutionary stages of the host star\cc{, as in the previous figures}. In the top panel, the pink solid and dashed lines represent the stellar radius and Roche limit, respectively.}
    \label{fig:temporal_evolution}
\end{figure}

\begin{figure*}[]
    \centering
     \includegraphics[width=0.85\hsize]{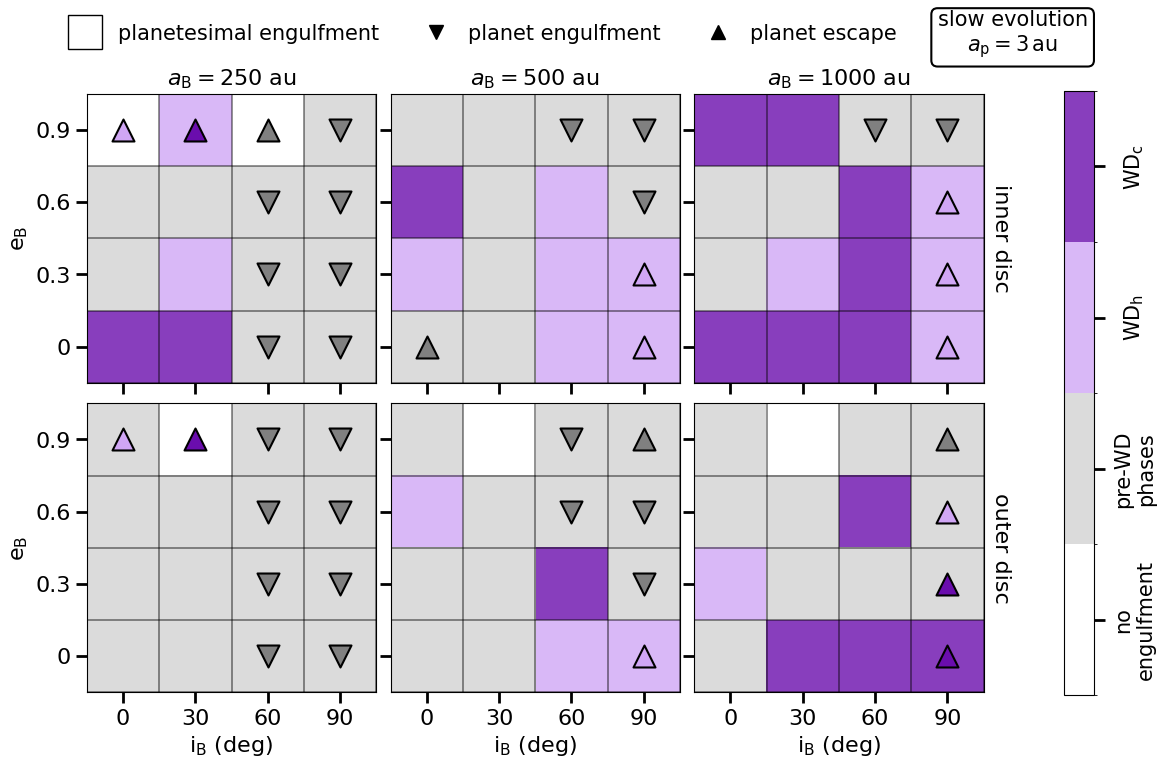}
    \caption{Simulation outcomes for the slow stellar-evolution model. The x- and y-axes correspond to the binary inclination $i_{\rm B}$ and eccentricity $e_{\rm B}$, respectively. From left to right, the columns correspond to binary semi-major axes \cc{$a_{\rm B}$=250, 500, and 1000 au. The top (bottom) panels correspond to planetesimal discs located interior (exterior) to the planet's orbit. In all simulations, the planet is initially located at $a_{\rm p} = 3$ au. Each grid cell represents a different initial binary configuration and indicates the outcome of the simulation. The background colour indicates planetesimal engulfment, while inverted and upright triangles denote planet engulfment and planet escape, respectively. The colour indicates the evolutionary stage of the host star at the time of each event: pre-WD (gray), WD$_{\rm h}$ (light purple), and WD$_{\rm c}$ (dark purple). White cells correspond to cases in which no planetesimals are engulfed, some survive, while others are ejected from the system.} 
    }
    \label{fig:slow_3auplanet}
\end{figure*}

\begin{figure}
\centering
\includegraphics[width=\hsize]{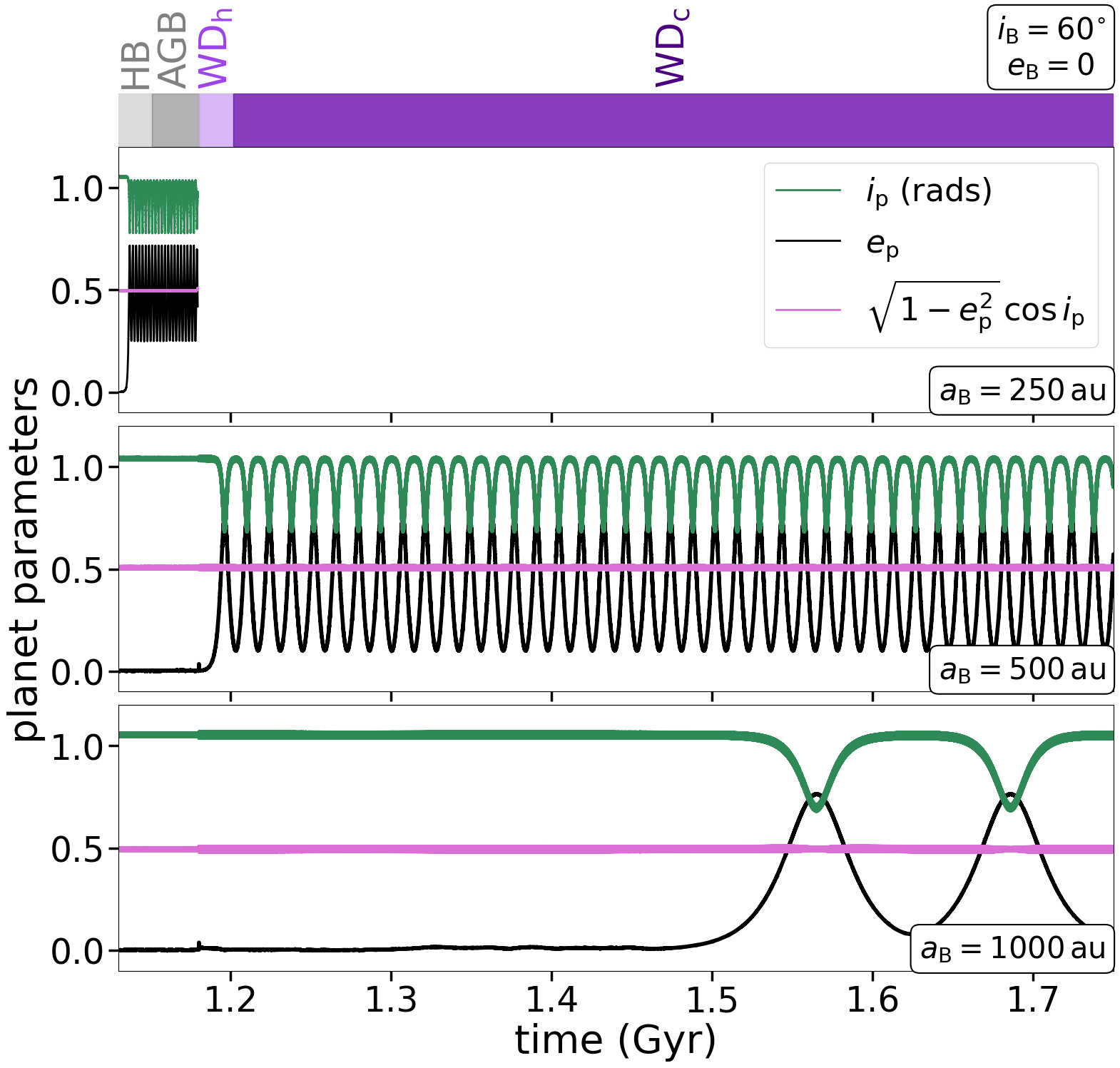}
   \caption{Temporal evolution of the planetary inclination (green), eccentricity (black), and the $z$-component of the angular momentum $j_{z}$ (pink), \cc{with time corresponding to the age of the host star. The planet is initially located at $a_{\rm p} = 3$~au and is perturbed by a binary with $i_{\rm B} = 60^\circ$ and $e_{\rm B} = 0$.} From top to bottom, the panels correspond to $a_{\rm B} = 250$~au, $a_{\rm B} = 500$~au, and $a_{\rm B} = 1000$~au, respectively.}
   \label{fig:kozai_examples}
\end{figure}

\begin{figure*}
    \centering
    \includegraphics[width=0.85\hsize]{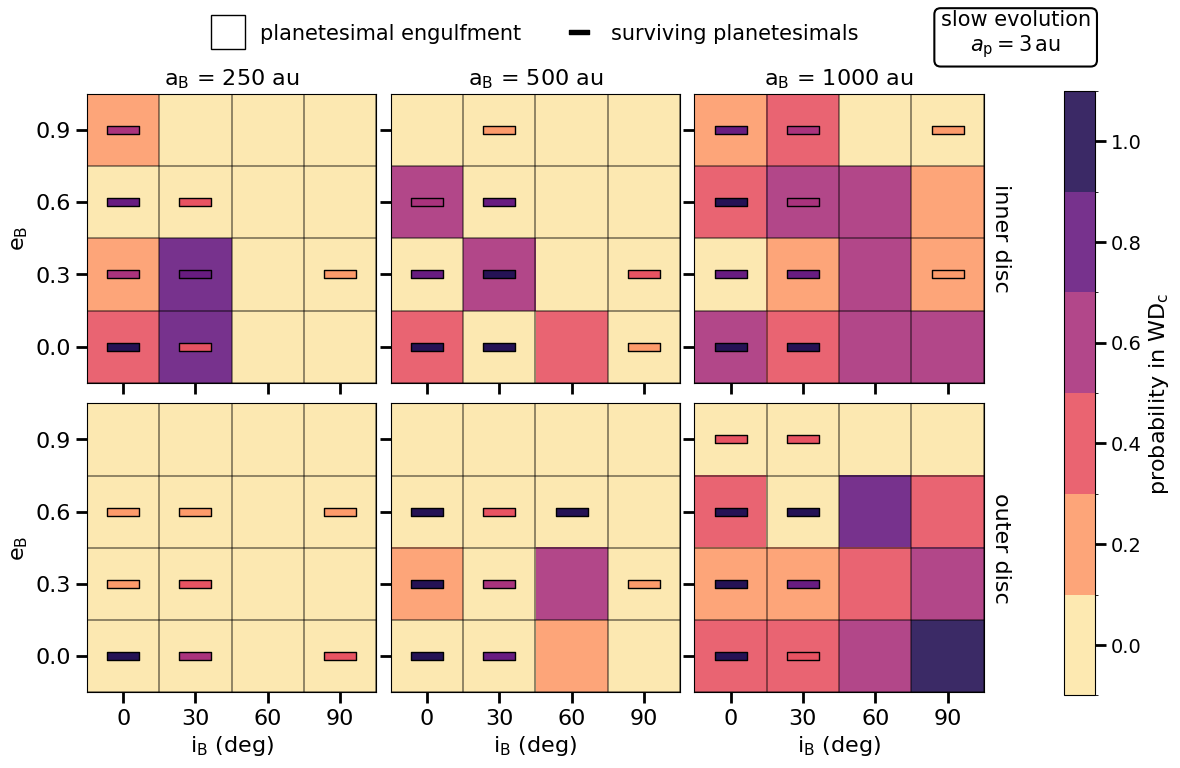}
    \caption{\cc{Probability of planetesimal engulfment during the WD$\rm_ c$ phase over five simulations} for a slowly evolving star. The panel layout is the same as in Fig.~\ref{fig:slow_3auplanet}. \cc{Background colours indicate the fraction of planetesimals engulfed by the host star during the WD$_{\rm c}$ phase, with darker colours corresponding to higher engulfment probabilities.} Horizontal lines mark configurations in which planetesimals survive until the end of the simulation.}
    \label{fig:probability_slow}
\end{figure*}

\begin{figure*}
    \centering
    \includegraphics[width=0.85\hsize]{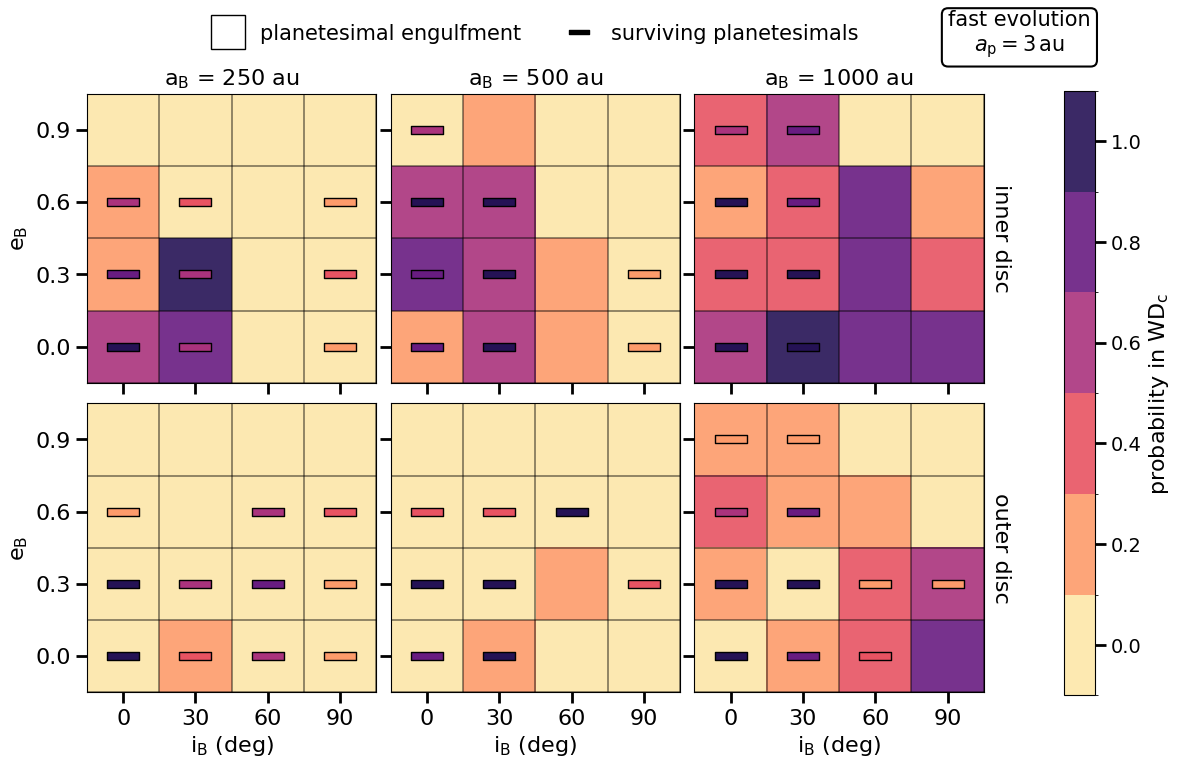}
    \caption{Planetesimal engulfment probability during the WD$_{\rm c}$ phase over five simulations, considering a fast evolution model for the host star. The panel layout, colour coding, and symbols are the same as in  Fig.~\ref{fig:probability_slow}.}
    \label{fig:probability_fast}
\end{figure*}

As already mentioned in Sect.~\ref{sec:Methods}, our setup consists of a host star orbited by a Jupiter-like planet and a planetesimal disc, together with a binary stellar companion. In our simulations, $t=0$ is defined at the Zero-Age Main Sequence (ZAMS), which we adopt as the reference point for consistency with the stellar-evolution models computed with MESA. The temporal evolution of all orbiting bodies in the system is shown in Fig.~\ref{fig:temporal_evolution}, with the pericenter distances (top panel) and eccentricities (bottom panel) as a function of the age of the host star. For clarity, we only show the evolutionary interval of interest, from the end of the horizontal branch (HB) phase until the planetesimal disc is fully depleted. The figure illustrates a representative configuration consisting of a slowly evolving host star, a binary (shown as a yellow line) at $a_{\rm B} = 1000$~au with $i_{\rm B} = 60^{\circ}$ and $e_{\rm B} = 0$, and a planet (black line) initially located at $a_{\rm p} = 3$~au exterior to an inner planetesimal disc composed of 25 planetesimals (shown in gray). The colour bar at the top indicates the evolutionary stage of the host star. To better illustrate the dynamical behaviour of the planetesimal disc, we do not show the full temporal evolution, which extends slightly beyond 5~Gyr in this particular case because the planet survives until the end of the simulation.

The top panel shows the pericenter evolution of all orbiting bodies and highlights key limits relevant to this study. The solid pink line marks the radius of the host star ($R_{\star}$), while the dashed pink line indicates the Roche radius \citep[$a_{\rm R}$,][]{1999Murray}, i.e., the minimum distance at which a planetesimal or planet can orbit the host star without being tidally disrupted. Both quantities evolve throughout the stellar lifetime: the stellar radius increases during the giant phases and contracts after WD formation, causing the Roche radius to vary accordingly. The bottom panel of Fig.~\ref{fig:temporal_evolution} shows the eccentricity evolution of all orbiting bodies.

During the pre-WD phases, the Roche limit lies within the stellar radius. As the star evolves and its radius decreases, however, the Roche limit moves outside the stellar surface. This transition marks an important change in the engulfment process: during the pre-WD phases, engulfment is governed by $R_{\star}$ \citep{2011ApJ...737...66K,2013ApJ...777L..30A}, whereas during the WD phase bodies approaching the star are expected to undergo tidal disruption upon crossing the Roche limit. Although the exact distance at which a body becomes irreversibly accreted during the WD phase remains uncertain because of the complex physics of tidal disruption and dissipation \citep{2016Veras}, throughout this work we adopt $R_{\star}$ and $a_{\rm R}$ as the engulfment boundaries during the pre-WD and WD phases, respectively.

Rocky material may be delivered onto the WD through the tidal disruption of minor bodies or more massive planets that cross the Roche limit \citep{2016Farihi}. In our simulations, however, bodies reaching the Roche limit are simply classified as engulfed, and the subsequent tidal disruption and debris evolution are not explicitly modeled. We therefore focus on identifying the system configurations capable of delivering remnant rocky bodies on to the WD, potentially leading to observable atmospheric pollution.

As expected from adiabatic orbital expansion driven by stellar mass-loss, which predicts a maximum expansion factor of $M_{\rm MS}/M_{\rm WD}$ \citep{1924MNRAS..85....2J, 2002Burleigh}, both the planet and the binary companion expand by factors of $\approx 2-3$ in all simulations. The fate of the inner planetesimals is instead determined by the competition between orbital expansion and stellar growth: for some bodies, the stellar radius overtakes their orbits before substantial expansion can occur, leading to engulfment.

During the WD stage, planetesimal ejections continue to occur as dynamical perturbations maintain high eccentricities within the system. As illustrated in Fig.~\ref{fig:temporal_evolution}, the planet can undergo eccentricity oscillations driven by ZKL cycles induced by the highly inclined binary companion, while its semi-major axis remains nearly constant. These oscillations modulate the planetary pericenter and contribute to the long-term dynamical excitation of the planetesimal disc. 

The efficiency of this mechanism depends on the architecture of the planetary system. We begin by examining our fiducial case corresponding to a planet located at $a_{\mathrm{p}} = 3$~au. Fig.~\ref{fig:slow_3auplanet} summarizes the outcomes of these simulations for the slow stellar-evolution model, showing a single run for each configuration. The outcomes reveal how the fate of the planet and the planetesimal discs depends on the binary inclination, eccentricity, and separation. The counterpart of the results for the fast stellar-evolution model is shown in Fig.~\ref{fig:fast_3auplanet}.

The top panels corresponds to configurations with the planetesimal disc located interior to the planet's orbit, whereas the bottom panels show systems with an exterior disc. From left to right, the columns show different initial binary semi-major axis: $a_{\mathrm{B}} = 250$ (left), $a_{\mathrm{B}} = 500$ (middle), and $a_{\mathrm{B}} = 1000$~au (right). These values correspond to the pre-AGB orbital configurations. As shown in Fig.~\ref{fig:temporal_evolution}, the binary orbit expands as the host star loses mass during its post-MS evolution, leading to larger binary separations by the time the system reaches the WD phase.

The colour scheme indicates the stellar evolutionary stage at which the event occurs: gray corresponds to the pre-WD phases, light purple to the WD$_{\rm h}$ phase, and dark purple to the WD$_{\rm c}$ phase. In each cell, the background colour records the last planetesimal engulfment event. White cells denotes the absence of engulfment; in most cases this means that planetesimals escape, although a small fraction of simulations have planetesimals surviving until the end. Planet-related outcomes are also shown, inverted triangles mark planet engulfments and upright triangles indicate escapes. 

Planetesimal engulfment occurs throughout the stellar evolution, from the HB to the WD$_{\rm c}$ phases. However, engulfment during the giant phases is not reflected in the atmospheric composition of the final WD. We therefore focus primarily on the dark-purple cells, which correspond to planetesimal accretion during the WD$_{\rm c}$ phase and thus represent the most relevant outcomes for observable WD pollution.

It is worth mentioning that in all configurations, the planetesimals are initially located within the Laplace radius \citep[$r_{\rm L}$,][]{2009AJ....137.3706T} and are therefore dynamically dominated by the gravitational potential of the planet, rather than by the binary companion. In this regime ($a < r_{\rm L}$), the planet acts as a dynamical shield, preventing secular perturbations from the inclined binary companion from driving ZKL oscillations within the planetesimal disc \citep{2017Petrovich}.
If the planet is later removed from the system, either through engulfment by the aging host star or through escape, the Laplace radius effectively shrinks to $r_{\rm L} \rightarrow 0$. The planetesimal disc then becomes directly exposed to secular perturbations from the binary, which can excite eccentricities and drive low-pericenter passages.

Planetesimal accretion during the WD$_{\rm c}$ phase is found across the full range of binary separations explored. However, its incidence increases from $a_{\rm B}=250$ to 500 and 1000 au, indicating that delayed pollution becomes progressively more common in wider binary configurations. Conversely, at smaller separations, planetesimal engulfment occurs more frequently during earlier evolutionary stages. This trend suggests that wider binary configurations favour delayed delivery of planetesimals to the host star, although such events are not exclusive to them.

In addition, planetesimal accretion during the WD$_{\rm c}$ phase is more frequent when the planetesimal disc is located interior to the planet's orbit. In this configuration, the smaller orbital distances increase the impact of dynamical perturbations from both the planet and the binary, facilitating the delivery of planetesimals onto star-grazing orbits \citep{Frewen_Hansen_2014}, thereby increasing the rate of accretion by the host star during the WD$_{\rm c}$ phase. By contrast, planetesimals in outer discs are more readily removed through ejections, as perturbations from both the planet and the binary companion can drive them onto unstable or escaping trajectories before they attain sufficiently small pericenter distances.

To understand the origin of the planetesimal accretion patterns shown in Fig.~\ref{fig:slow_3auplanet}, it is necessary to examine the evolution of the planet itself, which acts as a dynamical link  between the binary and the planetesimal disc. As mentioned in Sect.~\ref{sec:Methods}, different regimes arise depending on the binary inclination. 

For low binary inclinations ($i_{\rm B}<40^\circ$), ZKL oscillations are not present and the planet remains relatively stable. In these systems, the main drivers of planetesimal delivery are stellar mass-loss and the resulting orbital expansion of both the planet and the planetesimal disc. The changing orbital architecture can enhance planet--planetesimal interactions and scattering events, driving some bodies onto WD-grazing orbits \citep{2011Bonsor}. Additional secular perturbations from the binary may further contribute to the long-term excitation of eccentricities, even in nearly coplanar configurations. 

For higher inclinations ($i_{\rm B}\ge40^\circ$), the planet enters the ZKL regime. In these cases, the binary acts as an external perturber, driving coupled oscillations of the planet's eccentricity and inclination, while conserving $j_{z} = \sqrt{1 - e_{\mathrm{p}}^{2}}\, \cos i_{\mathrm{p}}$, the $z$-component of the angular momentum. As a result, the planet's eccentricity can get excited to large values, reducing the pericenter distance and potentially leading to tidal interactions with the host star, planetary engulfment, or even ejection from the system.

\citet{2017Petrovich} showed that, once the planet no longer shields the planetesimal disc, the binary can efficiently excite planetesimals through ZKL oscillations. In their study, the planet was removed artificially to mimic stellar engulfment and isolate the subsequent evolution of the planetesimal belt. Our simulations recover the same dynamical mechanism, but show that the timing at which the planetesimals become exposed to the binary perturbation depends on the prior dynamical evolution of the planet. In particular, wider binaries delay the onset of ZKL oscillations acting on the planet, extending the interval over which the planet continues to shield the planetesimal population.

To illustrate this behaviour and its dependence on binary separation, Fig.~\ref{fig:kozai_examples} presents the dynamical evolution of three representative systems in the ZKL regime, with $a_{\rm B}$ initially at 250, 500, and 1000~au (top, middle, and bottom panels, respectively). The binary inclination and eccentricity are fixed at $i_{\rm B} = 60^{\circ}$ and $e_{\rm B} = 0$. The figure shows the temporal evolution of the eccentricity $e_{\rm p}$ (black), inclination $i_{\rm p}$ (green), and $j_{z}$ (pink), with all quantities shown in the reference frame defined by the host star--binary companion plane. As expected for ZKL oscillations, the planet's eccentricity and inclination are anti-correlated, with eccentricity maxima corresponding to inclination minima (and vice versa), while $j_{z}$ remains nearly constant.

When the binary is closer ($a_{\rm B} = 250$~au), ZKL oscillations are triggered during the pre-WD phases, leading to planetary engulfment during the AGB phase through the rapid excitation of the planet's eccentricity. In contrast, for wider binaries, the planet remains on a near circular orbit and can survive the AGB phase, while the onset of ZKL oscillations is progressively delayed as the binary separation increases. For $a_{\rm B} = 500$~au, the oscillations are triggered during the WD$_{\rm h}$ phase, whereas for $a_{\rm B} = 1000$~au they only appear during the later WD$_{\rm c}$ phase.

The maximum eccentricity attained in all three systems is $e_{\rm p}\approx 0.75$, consistent with the classical quadrupole-level ZKL prediction for an initially circular orbit with $i_{\rm B}=60^{\circ}$ \citep{1997Natur.386..254H,2000ApJ...535..385F}. As expected from the secular theory, increasing $a_{\rm B}$ increases the characteristic timescale of the ZKL oscillations, while the maximum eccentricity reached by the planet remains nearly unchanged.
This delay in the onset of ZKL oscillations has important consequences for WD pollution. In systems with small binary separations, the dynamical instability is triggered before the host star becomes a WD, often leading to early planetesimal accretion and, in some cases, planetary removal. As a result, little material remains available to pollute the WD during its cooling phase. In contrast, wider binaries experience the onset of ZKL oscillations during the WD stage itself, allowing both the planet and the planetesimal disc to survive for longer timescales and promoting planetesimal accretion during the WD$_{\rm c}$ phase. This trend is also reflected by comparing Figs.~\ref{fig:slow_3auplanet} and \ref{fig:kozai_examples}. For the representative case with $i_{\rm B}=60^\circ$ and $e_{\rm B}=0$, the background colour in Fig.~\ref{fig:slow_3auplanet} changes from gray to progressively darker shades of purple as the binary separation increases, indicating that planetesimal accretion occurs at progressively later times, in agreement with the delayed onset of the planet's ZKL oscillations shown in Fig.~\ref{fig:kozai_examples}. This behaviour is also consistent with the observed prevalence of polluted WDs in wider binary systems, as shown in Fig.~\ref{fig:evol_track_observations}.

As in the case of planetesimals, binary separation has a significant impact on planetary survival. Systems with smaller values of $a_{\rm B}$ are more likely to lose their planets. For the closest binaries ($a_{\rm B}=250$ au), planetary removal occurs over a broad range of binary configurations. High-inclination systems tend to lose the planet during the HB phase as a consequence of strong ZKL-driven eccentricity excitation, while highly eccentric binaries can also lead to planetary ejection, although typically only near the end of the simulation. As the binary separation increases, the dependence on eccentricity becomes weaker and planetary survival is primarily controlled by the binary inclination. In particular, systems with $i_{\rm B}=90^\circ$ mostly result in planetary ejections. Although planetary removal during the AGB phase is often driven by tidal interactions with the expanding host star, perturbations from the binary can substantially modify the planet's orbital evolution, as illustrated by the ZKL-driven evolution described above. The dependence of planet survival on binary separation, inclination, and eccentricity demonstrates that the companion plays an important role in determining whether a planet ultimately survives, is engulfed by the evolving star, or is ejected from the system.

In the highest-inclination configurations ($i_{\rm B}=90^\circ$), where ZKL oscillations drive the planetary eccentricity to extreme values, short-range forces become increasingly important. Such extreme eccentricity excitation is characteristic of the high-eccentricity migration regime originally proposed for Hot Jupiter formation \citep[e.g.,][]{Wu2007,2007ApJ...669.1298F,2015ApJ...799...27P} and later extended to planets orbiting WDs \citep{2020ApJ...904L...3M}. Although tidal dissipation is included in our integrations, general relativistic (GR) precession is not. GR is expected to limit the maximum eccentricity attained during each ZKL cycle \citep{2015MNRAS.447..747L,Munoz2016}. Rather than preventing migration altogether, relativistic precession can stabilize the secular evolution by suppressing the most extreme eccentricity excursions while still allowing the periapsis to become sufficiently small for tidal dissipation to operate efficiently. Consequently, some systems that experience planetary escape in our pure N-body integrations may instead undergo high-eccentricity tidal migration, ultimately leading to orbital decay and engulfment, as shown for WD 1856+534 \citep{2020ApJ...904L...3M}.

The omission of GR is expected to have the largest impact in the widest binary configurations ($a_{\rm B}=500$--$1000$ au), where the secular forcing is weaker and therefore more susceptible to suppression by relativistic precession. In these systems, the combination of GR and tidal dissipation may favour high-eccentricity migration over the extreme eccentricity excursions that lead to planetary escape in our simulations, potentially reinforcing the wider binary pollution trend.

\begin{figure*}
\centering
\includegraphics[width=0.85\hsize]{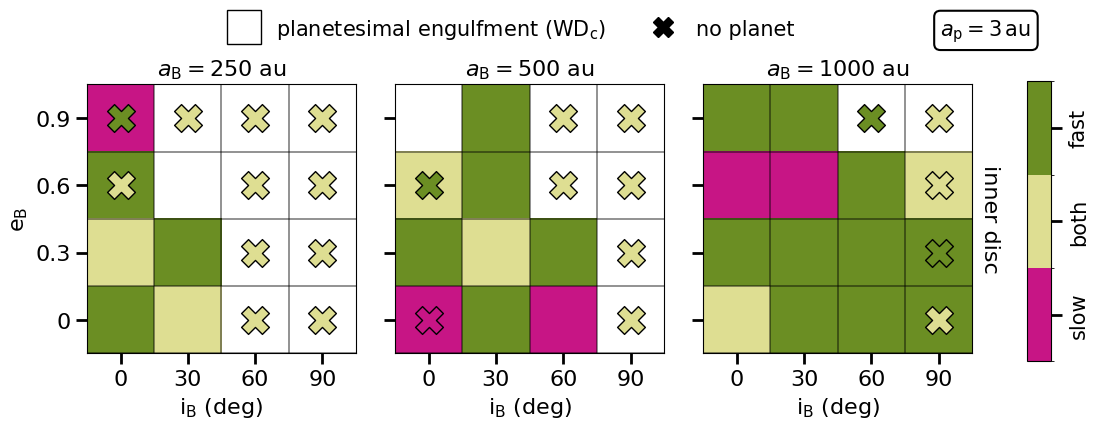}
   \caption{Comparison of planetesimal engulfment during the WD$_{\rm c}$ phase between the slow and fast evolution models for a planet located at $a_{\rm p}=3$ au. The background colour identifies whether planetesimal engulfment occurs in the fast (green), slow (pink), or in both evolution models (light yellow). White cells correspond to systems in which no planetesimals are engulfed during the WD$_{\rm c}$ phase in either model. The cross symbols mark configurations that become planet-free due to planet engulfment or ejection at any stage of system's evolution. Only configurations with an inner planetesimal disc are shown.} 
    \label{fig:comparasion_stellar_evolution}
\end{figure*}

To further investigate this configuration ($a_{\rm p}=3$~au) in detail, we perform five independent simulations per setup, to account for the chaotic nature of the dynamics. Although this sample size is insufficient for a rigorous statistical characterization, it is adequate to identify robust qualitative trends. To assess how frequent these engulfment events are during the WD$_{\rm c}$ phase, Fig.~\ref{fig:probability_slow} shows the same configuration with the planet initially located at $a_{\mathrm{p}} = 3$~au and a slowly evolving host star, but focuses on the probability of WD pollution during the cold stage. The panel layout follows the same scheme as in Fig.~\ref{fig:slow_3auplanet}. The background colour of each cell indicates the probability of planetesimal engulfment during the WD$_{\rm c}$ phase, computed from the five independent realizations performed for each configuration. The central horizontal bar in each cell encodes the probability that at least one planetesimal survives until the end of the simulation. In these cases, the planetesimal disc is not completely depleted, implying that rocky material remains available for potential pollution. 

In this probability plot, we recover the same qualitative trends identified in Fig.~\ref{fig:slow_3auplanet}: (i) inner planetesimal discs exhibit a higher probability of engulfment during the WD$_{\rm c}$ phase, whereas outer discs are more efficiently depleted through ejections; and (ii) WD$_{\rm c}$ pollution becomes more frequent as the binary separation increases, independent of whether the planetesimal disc is located interior or exterior to the planet's orbit.

While the individual simulations shown in Fig.~\ref{fig:slow_3auplanet} already suggested this behaviour, the probability map reveals more clearly that WD$_{\rm c}$ pollution is strongly suppressed in systems with highly eccentric companions ($e_{\rm B} = 0.9$), regardless of binary separation. In these cases, most planetesimals are engulfed or ejected during earlier evolutionary stages, leaving little remaining material to pollute the WD.

The increase in WD$_{\rm c}$ pollution toward wider binary separations is consistent with the observational trends shown in Fig.~\ref{fig:evol_track_observations} from the PEWDD database, where most polluted WDs are found in systems with binary separations larger than $1000$~au. \cc{A recent analysis of the Dark Energy Spectroscopic Instrument Data Release 1 (DESI DR1) also reports a lower metal-enrichment fraction at smaller binary separations and suggests that binary systems may experience more rapid depletion of their planetary material \citep{2026Xu}.} This agreement suggests that close binary companions may destabilize planetary systems too early, depleting the reservoir of small bodies before the host star reaches the WD phase. Moreover, even in configurations where ZKL oscillations dominate the secular evolution, stellar mass-loss may further contribute to the destabilization of the planetesimal disc and enhance the delivery of material onto the WD.

As described in Sect.~\ref{sec:Methods}, the specifics of post-MS stellar evolution remain uncertain. Although different prescriptions are generally adopted for different evolutionary stages, such as the RGB \citep{1975Reimers} and AGB \citep{1995Blocker} phases, the corresponding efficiency parameters remain poorly constrained. Previous studies have also shown that stellar mass-loss can play an important role in triggering dynamical instabilities and enhancing the delivery of planetesimals onto WDs through scattering processes \citep{2011Bonsor}.

To assess the sensitivity of our results to the adopted stellar-evolution model, we thus consider two prescriptions with different mass-loss efficiencies: a slow evolution model (already discussed in Figs.~\ref{fig:slow_3auplanet}, \ref{fig:kozai_examples}, and \ref{fig:probability_slow}) and a fast evolution model. The latter reaches the WD phase approximately 0.01~Gyr earlier than the former. Although this difference is small compared to the total stellar lifetime, it corresponds to roughly $10^{6}$ planetary orbits in the system considered here and may therefore affect their long-term dynamical evolution.

Fig.~\ref{fig:probability_fast} shows the same quantities presented in Fig.~\ref{fig:probability_slow}, but for the fast evolution model. The overall behaviour is largely unchanged: pollution events remain more common in wider binaries and in systems where the planetesimal disc is located interior to the planet's orbit. We also find that planetesimal survival is higher in systems with low binary inclinations. Similarly, surviving planetesimals are more frequently found in the outer than in the inner disc.

To directly compare the occurrence of planetesimal engulfment during the WD$_{\rm c}$ phase, Fig.~\ref{fig:comparasion_stellar_evolution} presents the results for the fiducial configuration with an inner planetesimal disc, which exhibits the highest incidence of WD pollution. Cross symbols indicate simulations that lost the planet, irrespective of the evolutionary phase in which the event occurs. Colours identify the stellar-evolution model in which planetesimal engulfment takes place: green corresponds to the fast evolution model, pink to the slow evolution model, and light yellow to events that occur in both.

A larger number of planetesimals are engulfed when the host star evolves faster, particularly in systems with low binary inclinations, where ZKL oscillations are absent. In this regime, the dynamical evolution is primarily governed by orbital expansion associated with stellar mass-loss, making the outcome more sensitive to the adopted mass-loss prescription. The fast evolution model ($\eta_{\rm RGB}=0.35$ and $\eta_{\rm AGB}=1.0$) employs larger mass-loss efficiencies than the slow evolution model ($\eta_{\rm RGB}=0.2$ and $\eta_{\rm AGB}=0.5$), resulting in shorter mass-loss timescales. Although the evolution remains largely adiabatic, the faster mass-loss may promote stronger dynamical instabilities and increase the likelihood of planetesimals being driven onto star-grazing orbits during the WD$_{\rm c}$ phase \citep[see also][]{2002ApJ...572..556D}.

The same happens at higher binary inclinations. In these systems, however, the dynamical evolution is dominated by secular perturbations from the stellar companion, particularly through the excitation of ZKL oscillations. As a consequence, variations in the stellar mass-loss rate have only a minor effect compared to the much stronger influence of the binary companion. This indicates that, once ZKL oscillations are activated, they become the primary mechanism driving planetesimal eccentricity growth and subsequent accretion onto the WD. Moreover, in the high-inclination configurations, the ZKL oscillations are initially excited in the planet and only later transmitted to the planetesimal disc. This delayed propagation of secular perturbations allows a larger fraction of the planetesimals to survive relatively undisturbed until the WD phase, increasing the reservoir of material available for late-time accretion.

In contrast, the survival of the planet itself shows little dependence on the adopted stellar-evolution model. Instead, it is primarily controlled by the binary configuration: systems with low binary inclinations almost always retain their planets, irrespective of the adopted stellar-evolution model. While a small number of configurations lead to planetary engulfment or escape in the fast but not in the slow evolution model, these differences are relatively minor. For systems with a planet at $a_{\rm p}=3$~au, planetary survival does not appear to be sensitive to the stellar mass-loss prescription. The most noticeable differences between the two models instead arise in the occurrence of planetesimal engulfment during the WD$_{\rm c}$ phase, particularly in low-inclination systems where secular perturbations from the binary companion remain weak. The few cases in which the planetary outcome differs between the two stellar-evolution models originate during the RGB/AGB evolution, when stellar expansion and mass loss determine whether the planet survives to the WD stage. This is broadly consistent with the results of \citet{2014ApJ...794....3V}, who found that planetary engulfment during the giant-branch evolution is only weakly sensitive to the adopted stellar mass-loss prescription. Within the parameter space explored here, however, the properties of the binary companion remain the primary driver of the system's long-term dynamical evolution and ultimate fate.

\subsection{\cc{Alternative configurations: $a_\mathrm{{p}}=2$ and $5$~au}}
\label{sect:2_5au}

    \begin{figure*}
       \centering
       \includegraphics[width=0.85\hsize]{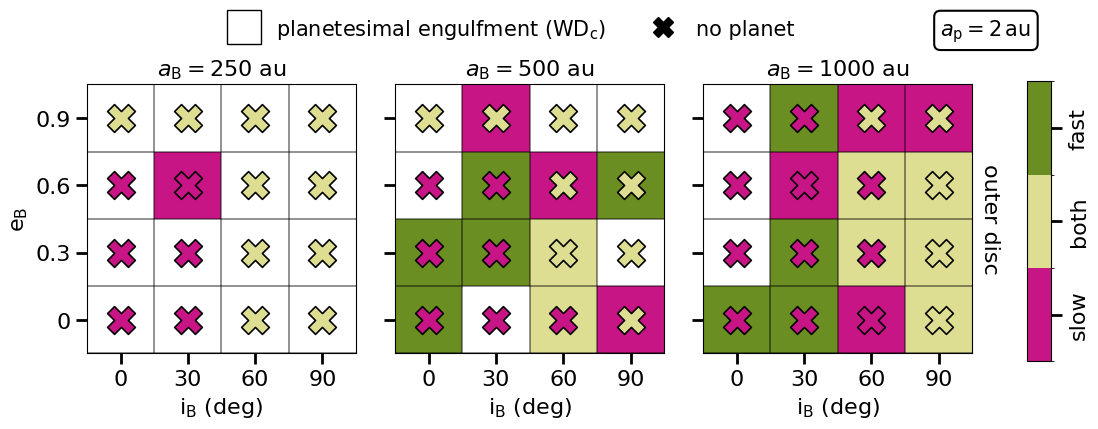}
       \caption{Comparison \cc{of planetesimal engulfment during the WD$_{\rm c}$ phase} between the slow and fast evolution stellar models for systems with $a_{\rm p}=2$~au and an outer planetesimal disc. Colours and symbols are defined as in Fig.~\ref{fig:comparasion_stellar_evolution}.}
       \label{fig:2au_slow_vs_fast}
    \end{figure*}
    
    \begin{figure*}
       \centering
       \includegraphics[width=0.85\hsize]{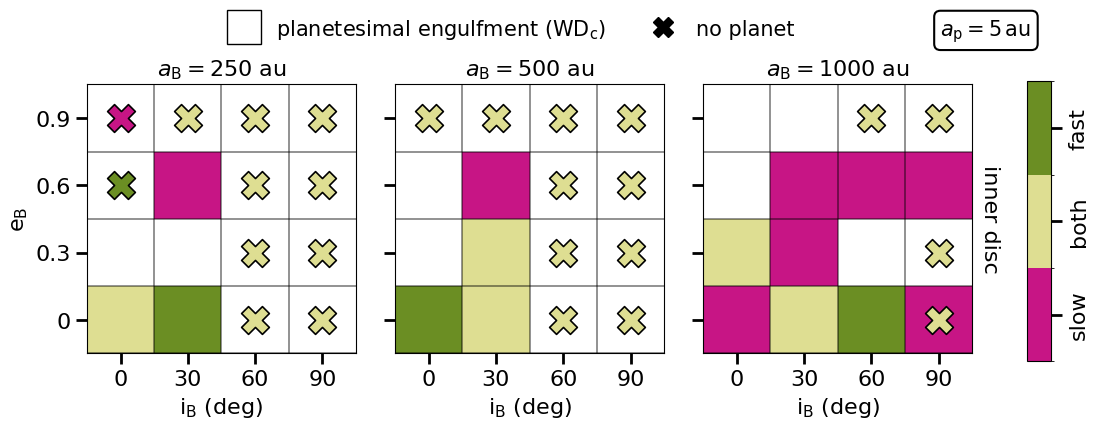}
       \caption{\cc{Comparison of planetesimal engulfment during the WD$_{\rm c}$ phase between the slow and fast stelar evolution models. The planet is located at $a_{\rm p}=5$~au, and the planetesimal disc is interior to its orbit. Colours and symbols are defined as in Fig.~\ref{fig:comparasion_stellar_evolution}.}} 
       \label{fig:5au_slow_vs_fast}
    \end{figure*}

    To assess the robustness of the trends identified in the fiducial configuration, we also performed simulations with the planet located at $a_{\rm p}=2$ and $5$~au for both stellar-evolution models. Because of the high computational cost of these integrations, only one simulation was performed for each configuration, unlike the fiducial $a_{\rm p}=3$~au case, for which multiple realizations were carried out to account for the chaotic nature of the dynamics. The results presented below should therefore be interpreted as illustrative rather than statistical.
    
    Fig.~\ref{fig:2au_slow_vs_fast} compares the planetesimal engulfment events during the WD$_{\rm c}$ phase between the slow and fast evolution models for configurations with the planet located at $a_{\rm p}=2$~au. In this case, the inner planetesimal disc spans $a \in [1.25, 1.75]$~au, while the outer disc extends over $a \in [2.5, 4]$~au. This figure is constructed in the same way as Fig.~\ref{fig:comparasion_stellar_evolution}. Results for the inner planetesimal disc are not shown because no planetesimal engulfment during the WD$_{\rm c}$ phase is observed in any of these cases. Instead, the planetesimal population is entirely depleted during the AGB phase as a consequence of the stellar radius expansion, to which the close-in planetesimals are particularly vulnerable. 
    
    As a result, all polluted systems in the $a_{\rm p}=2$~au configuration are associated with the outer planetesimal disc. This contrasts with the fiducial $a_{\rm p}=3$~au case, where WD pollution occurs predominantly in systems with an inner disc. A possible explanation is that, for $a_{\rm p}=2$~au, the outer planetesimal disc is initially located closer to the host star than in the other planetary configurations. As the host star loses mass and the planetary orbits expand, enhancing scattering and increasing the probability of bodies reaching low-pericenter orbits around the WD. Moreover, the outer planetesimal disc survives in most configurations, leaving a substantial reservoir of material that could continue to supply planetesimals onto the WD beyond the timespan explored in our simulations.
    
    For the $a_{\rm p}=2$~au configuration, the two stellar-evolution models produce a comparable number of planetesimal engulfment events during the WD$_{\rm c}$ phase. The main difference is that the slow evolution model retains surviving planetesimals in a larger number of configurations, with a significant fraction of planetesimal discs remaining only partially depleted by the end of the simulations. Given that only one simulation was performed for each configuration, however, this difference may simply reflect the stochastic nature of the N-body evolution rather than a systematic dependence on the adopted stellar-evolution prescription. As in the fiducial case, systems with $a{\rm B}=250$~au show an almost complete absence of planetesimal engulfment during the WD$_{\rm c}$ phase in both stellar-evolution models, with only a single event occurring in the slow evolution model. As the binary separation increases, the number of polluted systems also increases, consistent with the trend found in the fiducial case.
    
    While planetesimal engulfment during the WD$_{\rm c}$ phase shows little dependence on the adopted stellar-evolution model, planetary survival is considerably more sensitive to the adopted prescription. For low binary inclinations, the planet survives the giant phases in the fast evolution model, unlike the outcome reported by \citet{2012Mustill}. This stronger dependence is expected because the planet lies closer to the host star, where tidal interactions are stronger \citep{2009Villaver} and therefore more sensitive to differences in the stellar-radius evolution and mass-loss history. In addition, before reaching the WD phase, the planet completes approximately $4\times10^5$ more orbits in the slow than in the fast evolution model, potentially allowing tidal effects to accumulate over more orbital cycles, thereby increasing the likelihood of planetary engulfment.
    
    These results can be compared with the findings of \citet{2014ApJ...794....3V}, who found that planetary engulfment during the RGB is only weakly dependent on the adopted stellar mass-loss prescription. However, the critical survival distance of a Jupiter-like planet changes slightly with the assumed mass-loss rate (see their Table~1). Although these are minor differences, the different survival rates found in our fast and slow models may reflect that the initial planetary orbit lies close to the critical survival distance, where relatively small changes in the stellar evolution can alter the final outcome. Moreover, while \citet{2014ApJ...794....3V} focused primarily on the RGB evolution, our simulations focus on the AGB phase, where mass loss is substantially stronger \citep{1993Vassiliadis} and the differences between the two stellar-evolution models become more pronounced.
    
    Fig.~\ref{fig:5au_slow_vs_fast} illustrates planetesimal engulfment during the WD$_{\rm c}$ phase for both the slow and fast evolution models, for a planet located at $a_{\rm p}=5$~au. In this setup, the inner planetesimal disc extends over $a \in [3.5, 4.5]$~au, whereas the outer disc spans $a \in [5.5, 7]$~au. As in the fiducial case, planetesimal accretion during the WD$_{\rm c}$ phase occurs more frequently in systems with an inner disc, because planetesimals scattered from the inner disc can more rapidly attain low-pericenter orbits and are eventually accreted onto the host star. By contrast, planetesimals originating from the outer disc are more likely to be ejected before reaching star-grazing trajectories.
    
    When $a_{\rm p}=5$~au, the incidence of WD pollution from planetesimals is broadly similar in the fast and slow evolution models for binary separations of $a_{\rm B}=250$ and $500$~au. A noticeable difference emerges only for the widest binaries ($a_{\rm B}=1000$~au), where the slow evolution model produces a higher incidence of planetesimal accretion onto the WD. Likewise, the number of configurations retaining surviving planetesimals is comparable in the two models, with no systematic trend favoring either prescription. In contrast, no difference is found in the survival of the planet itself.
    At this orbital distance, both the planet and the planetesimal disc are less affected by stellar tides and mass-loss than in the $a_{\rm p}=2$ and $3$~au configurations. Consequently, the dynamical evolution is dominated by perturbations from the binary companion, rendering the overall outcomes largely insensitive to the adopted mass-loss parametrization.
    
    Overall, our results demonstrate that the survival limits derived by \citet{2012Mustill} can be significantly modified by the presence of a stellar companion. While their study predicts that Jupiter-like planets initially located at $a_{\rm p}=5$~au around a $2\,M_{\odot}$ star should survive the AGB phase, we find several configurations in which the planet is ultimately removed from the system. These outcomes occur primarily in systems with a  highly inclined binary, and in systems with highly eccentric binaries ($e_{\rm B}=0.9$), which further strengthen the secular perturbations. This comparison emphasizes that the long-term fate of planets around evolving stars is governed not only by stellar mass-loss and tidal interactions, but also by the secular perturbations induced by wide binary companions.

\section{Discussion}\label{sec:discussion}

The fate of the planetary system is governed by the competition between orbital expansion driven by stellar mass-loss, orbital decay due to stellar tides \citep[e.g.,][]{2026MauchSoriano}, and secular perturbations induced by the binary companion. During the AGB phase, stellar mass-loss tends to expand the planetary orbit, whereas tidal dissipation removes orbital energy and angular momentum. When the binary companion excites ZKL oscillations, the resulting eccentricity excitation can drive the planet to sufficiently small pericenter distances for tidal dissipation to become efficient, ultimately leading to engulfment and determining the amount of rocky material that survives into the WD phase.

The prevalence of WD pollution in systems with a wider binary can be understood in terms of the delayed propagation of secular perturbations from the binary companion to the planetesimal disc. Because ZKL oscillations are first excited in the planet, the planetesimals can remain comparatively undisturbed for an extended period before the perturbations propagate through the system. As a result, a larger reservoir of planetesimals survives into the WD phase, where subsequent eccentricity excitation can drive late-time accretion events. This provides a possible explanation for the observational trend shown in Fig.~\ref{fig:evol_track_observations}, where polluted WDs in binary systems are predominantly found at projected separations larger than $1000$~au \cc{\citep[see also][]{2026Xu}}.

\begin{figure}
\centering
\includegraphics[width=\columnwidth]{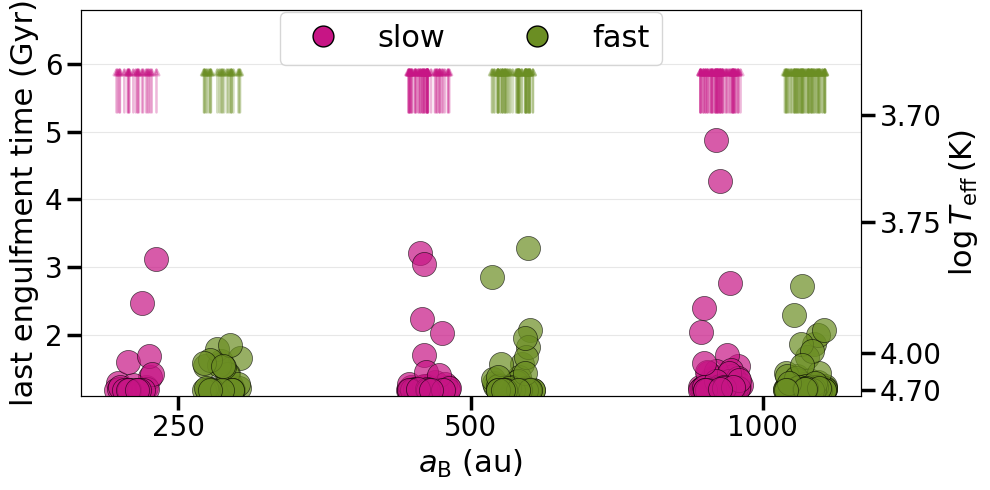}
   \caption{Time of the last planetesimal engulfment during the WD phase in each simulation. Pink and green dots correspond to the slow and fast stellar-evolution models, respectively. The left y-axis shows the stellar age, while the right y-axis indicates the WD effective temperature. The x-axis represents the binary separation, with a small horizontal offset applied to individual points for visualization only. Arrows denote simulations in which at least one planetesimal survives until the end of the integration.}
   \label{fig:max_engulfment_WD}
\end{figure}

A complementary observational constraint comes from the broad range of WD cooling ages over which atmospheric pollution is detected.
Because heavy elements rapidly diffuse out of WD atmospheres, the detection of atmospheric pollution implies recent or ongoing accretion \citep{2009Koester}. The broad temperatures distribution of polluted WDs (Fig.~\ref{fig:evol_track_observations}) therefore indicates that the delivery of rocky material must operate throughout the WD cooling sequence. To asses whether our simulations can reproduce this behaviour, Fig.~\ref{fig:max_engulfment_WD} shows the timing of the latest planetesimal engulfment events as a function of binary separation. The simulations span a similarly broad range of WD cooling ages, with wider binaries producing both the latest engulfment events and a larger fraction of systems retaining planetesimals at the end of the integrations (\cc{$T_{\rm eff}\approx 5000$~K}, upward arrows). The latter indicates that accretion may continue beyond the simulated interval, particularly in wider binaries, which can preserve long-lived planetesimal reservoirs capable of sustaining pollution over Gyr timescales.

This behaviour also support the picture proposed in previous studies, which argued that explaining pollution across the full range of WD cooling ages requires both a long-lived planetesimal reservoir and a sustained delivery mechanism \citep{2014MNRAS.439.3371W,2014Koester,2017Petrovich,2022MNRAS.513.4178O}. We note, however, that the engulfment times shown in Fig.~\ref{fig:max_engulfment_WD} correspond to the arrival of planetesimals at the WD Roche limit, rather than to the onset of atmospheric pollution. The subsequent tidal disruption of the planetesimals and the formation of an accretion disc occur on additional timescales, implying that these engulfment times should be regarded as lower limits on the onset of WD pollution.

Although the qualitative agreement between our simulations and the observed distribution of polluted WDs is encouraging, this comparison should be interpreted with caution \cc{as the binary fractions being compared are derived from samples selected using different criteria and therefore do not represent exactly the same quantity. In most cases, the binary fraction of polluted WDs is determined by first identifying WDs with atmospheric metal pollution and subsequently searching for binary companions. This differs from selecting a population of wide binary systems containing WDs and then search systematically for atmospheric metal pollution among their WD components. Thus, the two approaches involve different sample-selection procedures, which should be taken into account when interpreting comparisons between observations and simulations.}

A comparison with previous studies highlights the sensitivity of planetary survival to the adopted system architecture and stellar-evolution model. Differences in the adopted stellar-evolution prescription may alter the competition between orbital expansion due to mass-loss and orbital decay due to tides. Using stellar-evolutionary tracks based on the mass-loss prescription of \citet{1993Vassiliadis}, \citet{2012Mustill} found that a Jupiter-like planet orbiting a $2\,M_{\odot}$ star required an initial semi-major axis of \cc{$a_{\rm p}\approx$3.5~au} to survive the AGB phase \citep[see also][]{2013MNRAS.432..500N}. In contrast, our simulations show that a planet initially located at $a_{\rm p}=3$ au can survive in a subset of configurations. This difference likely reflects both the presence of a binary companion in our models and differences in the adopted stellar-evolution prescriptions. In our models, planetary evolution is influenced not only by stellar tides and mass-loss, but also by the presence of a binary companion, whose separation, inclination and eccentricity modify the secular evolution of the planet. Similarly, \citet{2017Veras} found that wider binaries favour planetary survival and that critical separation for stability of the system increases with binary eccentricity, considering coplanar planetary and binary orbits. Although their study considered a different planetary architecture, these results support the dependence of the stability boundary on the dynamical configuration of the system. These differences suggest that the critical distance for planetary survival is not universal, but depends on both the dynamical architecture of the system and the details of the stellar-evolution model.

Our simulations yield a higher level of WD accretion than previous single-planet studies. \citet{Frewen_Hansen_2014} showed that stellar mass-loss can widen resonances, trap planetesimals, and gradually increase their eccentricities until they reach the WD disruption radius, while \citet{2010ApJ...715.1036D} found that stellar evolution can also lead to resonant capture in the presence of wind drag. In the simulations of \citet{Frewen_Hansen_2014}, however, only a small fraction of particles were ultimately disrupted in their simulations. Part of this discrepancy may reflect differences in the adopted post-MS integration timespan, since resonant delivery can operate over extended timescales. In our case, although the planet is initially placed on a circular orbit, the presence of a binary companion can excite its eccentricity through ZLK oscillations in systems that lie within the corresponding dynamical regime. In this sense, our results may be consistent with the trend identified by \citet{2012ApJ...747..148D}, who found that higher planetary eccentricities enhance the delivery of small bodies onto the WD.

We considered two stellar-evolution models characterized by different mass-loss efficiency parameters, which lead to different mass-loss histories during the RGB and AGB phases. Determining which model more accurately reflects stellar evolution remains challenging, given the existing uncertainties in mass-loss prescriptions. For the \citet{1975Reimers} prescription, empirical calibrations based on old, metal-rich open clusters suggest values in the range $\eta_{\rm RGB}\approx0.1$--$0.3$ \citep{2012Miglio}. On the other hand, the efficiency parameter of the \citet{1995Blocker} prescription remains even more uncertain, with values spanning more than an order of magnitude in literature. While \citet{1995Blocker} originally recommended a value of $\eta_{\rm AGB}=1.0$, subsequent studies have favoured significantly lower efficiencies. For instance, \citet{2014Kalirai} explored values of $\eta_{\rm AGB}=0.2$ and $0.05$ by calibrating the models against the observed initial--final mass relation and constraints on third dredge-up for solar-metallicity stars. Other studies have adopted even lower efficiencies depending on the stellar-evolution code and calibration method; for example, \citet{2012Ventura} used $\eta_{\rm AGB}=0.02$ to reproduce observations of low-metallicity stars in the Large Magellanic Cloud. The uncertainty in AGB mass loss extends beyond the choice of $\eta_{\rm AGB}$, as the physical mechanism and their evolution during AGB phase remain poorly constrained, particularly during its final stages \citep{2026Cui}. This uncertainty motivates exploring different mass-loss histories when assessing their impact on post-MS planetary dynamics.

A limitation of our study is the small number of planetesimals included in each simulation. which was motivated by computational constraints. Therefore, we do not aim to derive accretion rates, as done by \citet{2011Bonsor} and \citet{2018Mustill}, or make quantitative predictions of the frequency of WD pollution. Instead, we focus on identifying dynamical configurations capable of delivering rocky material onto the WD during the WD$_{\rm c}$ phase, when atmospheric pollution is expected to result from external accretion. Although a larger number of particles would be required for statistically robust estimates, the delivery of planetesimals onto the WD serves as an indicator of the pollution potential of a given configuration. Additionally, our simulations do not include Galactic tides, which may become important for the widest binary configurations ($a_{\rm B}=1000~$~au) and could further influence the long-term evolution of the surviving planetesimal reservoirs.

\section{Conclusions}\label{sec:conclusion}

Polluted WDs provide compelling evidence that remnants of planetary systems can survive the post-MS evolution of their host stars and continue to deliver rocky material onto the WD. While several mechanisms have been proposed to drive this delivery, the role played by wide binary companions in combination with stellar evolution remains poorly explored. In this work, we used N-body simulations to investigate how stellar mass-loss, tidal interactions, and perturbations from a wide binary companion influence the long-term evolution of planetary systems and the subsequent accretion of planetesimals onto WDs. 

\cc{Our main findings can be summarized as follows:}
\cc{\begin{enumerate}
    \item The architecture of the wide binary is the dominant factor controlling the dynamical evolution and pollution outcome. Wider binaries delay the onset of ZKL oscillations, allowing planets and planetesimal reservoirs to survive longer and increasing the efficiency and duration of planetesimal delivery onto the white dwarf. This naturally leads to higher pollution rates for wider binary separations, consistent with the observed prevalence of polluted WDs in wide binaries.
    \item The survival and location of planetesimal reservoirs strongly influence the pollution history. Reservoirs located interior to the planet's orbit are more efficiently delivered onto the WD, whereas reservoirs located exterior to the planet's orbit are more prone to ejection. At the same time, long-lived reservoirs can sustain metal pollution over Gyr timescales, extending the pollution phase beyond the duration of our simulations.
    \item The adopted stellar-evolution prescription has a secondary effect on the dynamical evolution. Differences in the stellar mass-loss history primarily affect the timing and efficiency of planetesimal delivery, while the overall dynamical outcome is mainly determined by the configuration of the binary. In particular, the survival of the planet is only weakly affected by the adopted stellar-evolution model, except for planets initially located very close to the central star.
\end{enumerate}}

\cc{Overall, our results suggest that the architecture of wide binaries, rather than uncertainties in stellar evolution, is the dominant factor controlling the dynamical delivery of rocky material onto WDs. The combination of binary architecture and long-lived planetesimal reservoirs can produce prolonged delivery of rocky material, providing a natural mechanism for metal pollution over Gyr timescales. As a next step, we will compare the pollution rates obtained for binary and single-star systems to quantify the relative contribution of wide binaries to the observed population of polluted WDs. }

\begin{acknowledgements}
\cc{The authors thank the anonymous reviewer for their thoughtful comments, which have helped us improve the quality of the manuscript. }
C.N. and C.A.G. acknowledge support from Agencia Nacional de Investigación y Desarrollo (ANID) through FONDECYT Iniciación 11230741 and FONDECYT Regular 1262342. 
C.C. acknowledges support from ANID post-doctoral grant n$^\circ$3230283 and CAS-ANID (CASSACA) n$^\circ$250005.
\end{acknowledgements}

\bibliographystyle{aa}
\bibliography{citas}

\begin{appendix}

\section{Additional methodology details}
\label{appendix:methodology}

    This section provides further details on the methodology presented in the main text.

\subsection{Stellar evolution}

    Rotationally induced instabilities are included using the standard \texttt{MESA} prescriptions, with a reduced efficiency for compositional mixing $f_{c}=1/30$, following \citet{2000Heger}, and no viscous transport. During the AGB phase, we adopt an exponential overshooting prescription for convective boundary mixing with an efficiency parameter $f=0.014$.

    To model the stellar tides, we adopt the weak-friction, constant time-lag formalism introduced by \citet{1979Mignard} and secularly developed by \citet{1981Hut}, coupled with the convective damping prescription appropriate for evolved stars with extended convective envelopes \citep{1977Zahn, 1989Zahn}. In this framework, tidal dissipation is represented through a constant time lag between the perturbing potential and the stellar tidal bulge, defined by Eq.~8 in \citet{2022Baronett} as
    \begin{equation}
        \tau=\frac{2R^3}{GMt_{f}}, \qquad \text{where} \qquad 
        t_{f}= \left ( \frac{MR^2}{L}  \right ) ^{1/3}
        \label{eq:tau}
    \end{equation}
    is the convective friction time (Equation~7 in \citealp{1989Zahn}).

\subsection{N-body dynamics}

    The companion mass is fixed at $1\,M_{\odot}$, corresponding to a representative mass-ratio ($q$) for wide binary systems and producing significant dynamical perturbations on the planetary system. Observationally, main sequence wide binaries exhibit a broad mass-ratio distribution, typically extending over \cc{$q\,$$=\,$0.1--1} and weighted toward lower-mass companions \citep{2017Moe}.
    
    The planet orbital distances are chosen to span configurations ranging from systems in which the planet is likely to be engulfed during stellar evolution, to cases where it is expected to survive the host star's post-MS evolution. \citet{2012Mustill} showed that tidal interactions during the giant phases can engulf Jovian planets initially located at distances of up to \cc{3~au} around a $1\,M_{\odot}$ star and up to \cc{5~au} around a $5\,M_{\odot}$ star. Similarly, \citet{2020Ronco} found a critical distance of $2.7$~au separating engulfment from survival for a Jupiter-mass planet orbiting a Solar-type star during the RGB phase. Motivated by these results, we adopt $a_{\rm p}=2$, 3, and 5~au to probe configurations close to the engulfment boundary as well as systems where the planet is expected to survive stellar evolution. This also allows us to investigate how different stellar-evolution models affect the long-term dynamical evolution of the system and the subsequent accretion of planetesimals during the WD phase.

\section{\cc{Fast evolution for a planet at $a_{\rm p}=3$~au}}
\label{appendix:fast_3auplanet}
\begin{figure*}
    \centering
     \includegraphics[width=0.85\hsize]{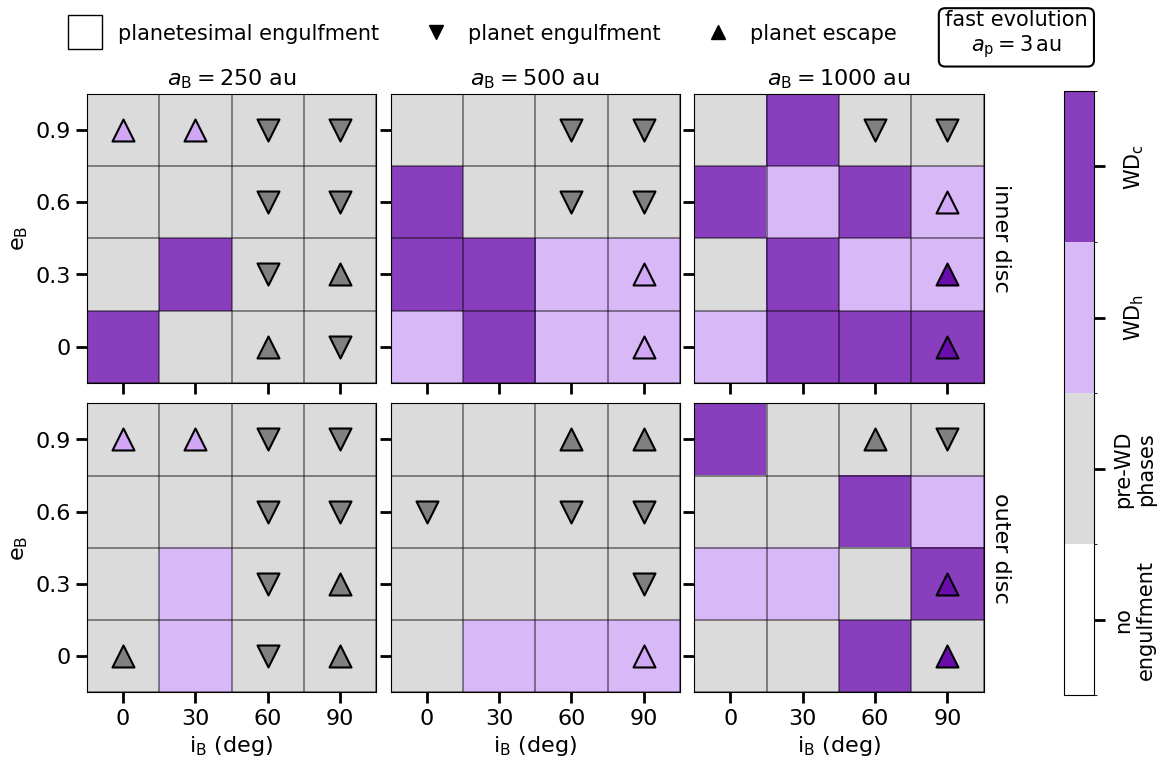}
    \caption{\cc{Simulation outcomes for the fast stellar-evolution model with a planet located at $a_{\rm p}=3$~au. The panel layout, colour coding, and symbols are the same as in Fig.~\ref{fig:slow_3auplanet}. 
    }}
    \label{fig:fast_3auplanet}
\end{figure*}

\cc{For completeness, Fig.~\ref{fig:fast_3auplanet} shows the corresponding results obtained with the fast stellar-evolution model. Overall, the results are broadly consistent with those presented in Fig.~\ref{fig:slow_3auplanet} for a slowly evolving star, with the main differences between the two evolutionary models discussed in Sect.~\ref{sec:results}.}

\end{appendix}

\end{document}